\documentclass[fleqn,usenatbib]{mnras}

\usepackage{newtxtext,newtxmath}

\usepackage[T1]{fontenc}
\usepackage{tabularx}
\usepackage{booktabs}

\allowdisplaybreaks

\DeclareRobustCommand{\VAN}[3]{#2}
\let\VANthebibliography\thebibliography
\def\thebibliography{\DeclareRobustCommand{\VAN}[3]{##3}\VANthebibliography}

\usepackage{graphicx}	

\def\m87{{M87$^{\ast} $}}

\newcommand{\ugas}{u_{\rm gas}}
\newcommand{\pgas}{p_{\rm gas}}
\newcommand{\adi}{\Gamma_{\rm gas}}

\newcommand{\ed}{\mathop{}\!\mathrm{d}}

\def\lsim{\mathrel{\raise.3ex\hbox{$<$\kern-.75em\lower1ex\hbox{$\sim$}}}}
\def\gsim{\mathrel{\raise.3ex\hbox{$>$\kern-.75em\lower1ex\hbox{$\sim$}}}}
\def\gtwid{\mathrel{\raise.3ex\hbox{$>$\kern-.75em\lower1ex\hbox{$\sim$}}}}
\def\proptwid{\mathrel{\raise.3ex\hbox{$\propto$\kern-.75em\lower1ex\hbox{$\sim$}}}}

\graphicspath{{./}{figures/}}

\title[Radiative Simulation Survey of M87*]{Survey of Radiative, Two-Temperature Magnetically Arrested Simulations of the Black Hole M87* I: Turbulent Electron Heating}

\author[A. Chael]{
Andrew Chael$^{1}$\thanks{E-mail: achael@princeton.edu}
\\
$^{1}$Princeton Gravity Initiative, Princeton University, Princeton NJ, 08540\\
}

\date{Accepted XXX. Received YYY; in original form ZZZ}

\pubyear{2025}

\begin{document}
\label{firstpage}
\pagerange{\pageref{firstpage}--\pageref{lastpage}}
\maketitle

\begin{abstract}
We present a set of eleven two-temperature, radiative, general relativistic magnetohydrodynamic (2TGRRMHD) simulations of the black hole M87* in the magnetically arrested (MAD) state, surveying different values of the black hole spin $a_*$. Our 3D simulations self-consistently evolve the temperatures of separate electron and ion populations under the effects of adiabatic compression/expansion, viscous heating, Coulomb coupling, and synchrotron, bremsstrahlung, and inverse Compton radiation.
We adopt a sub-grid heating prescription from gyrokinetic simulations of plasma turbulence.
Our simulations have accretion rates $\dot{M}=(0.5-1.5)\times10^{-6}\dot{M}_{\rm Edd}$ and radiative efficiencies $\epsilon_{\rm rad}=3-35\%$.
We compare our simulations to a fiducial set of otherwise identical single-fluid GRMHD simulations and find no significant changes in the outflow efficiency or black hole spindown parameter.
Our simulations produce an effective adiabatic index for the two-temperature plasma of $\adi\approx1.55$, larger than the $\adi=13/9$ value often adopted in single-fluid GRMHD simulations.
We find moderate ion-to-electron temperature ratios in the 230 GHz emitting region of $R=T_{\rm i}/T_{\rm e}{\approx}5$.
While total intensity 230 GHz images from our simulations are consistent with Event Horizon Telescope (EHT) results, our images have significantly more beam-scale linear polarization ($\langle|m|\rangle\approx 30\%$) than is observed in EHT images of M87* ($\langle|m|\rangle<10\%$).
We find a trend of the average linear polarization pitch angle $\angle\beta_2$ with black hole spin consistent with what is seen in single-fluid GRMHD simulations, and we provide a simple fitting function for $\angle\beta_2(a_*)$ motivated by the wind-up of magnetic field lines by black hole spin in the Blandford-Znajek mechanism.
\end{abstract}




\section{Introduction}
\label{sec:intro}
Most supermassive black holes are surrounded by hot, low-luminosity plasma accretion flows \citep{GreeneHo07, Ho08}.
Emission from these low-luminosity accretion flows is predominantly produced by
synchrotron radiation from relativistic electrons at millimetre wavelengths \citep{Ichimaru77,NarayanYi94,Yuan14}.
The Event Horizon Telescope (EHT) Very-Long-Baseline Interferometry (VLBI) experiment has resolved millimetre-wavelength synchrotron images on scales of the event horizon $r \lsim 5 \, r_{\rm g}=5\,GM/c^2$ in the nearby low-luminosity AGN M87* \citep{PaperIV} and the Galactic Centre black hole Sgr A* \citep{SgrAPaperIII}.

The elliptical galaxy M87 displays a prominent extragalactic jet \citep{Curtis1918} extending from the central supermassive black hole \citep{Hada11} to kiloparsec distances,
with emission across the electromagnetic spectrum \citep{EHT2024M87}.
Extragalactic jets like that in M87 are widely believed to be powered by the central black hole's spin energy, extracted by magnetic fields threading the event horizon \citep{BZ}, though an alternative power source from rotational energy in the accretion disc \citep{BP} cannot yet be ruled out.
Millimetre-wavelength VLBI images of M87* in full polarization map out the geometry and strength of the magnetic field just outside the event horizon and have the potential to directly probe the energy source of M87's extragalactic jet \citep{Chael23, Johnson23,MidRange}.

General Relativistic Magnetohydrodynamic (GRMHD) simulations are an essential tool for understanding the properties of hot accretion flows and jets around supermassive black holes \citep[e.g.][]{Komissarov99,Gammie03}. Using general relativistic ray-tracing and synchrotron radiation transfer codes \citep[e.g.][]{Dexter16,ipole,Prather23}, it is possible to generate simulated horizon-scale millimetre images of M87*, Sgr A*, and other sources \citep[e.g.][]{Zhang24} for direct comparison to VLBI observations.
An extensive comparison of simulated images from GRMHD simulations to polarized EHT observations indicates that M87* \citep{PaperVIII,PaperIX} and Sgr A* \citep{SgrAPaperVIII} are likely in the magnetically arrested (MAD) state of black hole accretion, where the magnetic field near the black hole is largely ordered and dynamically important \citep{Narayan03,Tchekhovskoy11}.

The hot accretion flows around M87* and Sgr A* are so low density that electrons and ions are not in thermal equilibrium \citep{Shapiro76,Rees82}.
Single-fluid GRMHD simulations assume equilibrium, and thus do not provide any direct constraints on the temperature of the electrons that produce the observed emission.
In practice when producing simulated GRMHD images, the ion-to-electron temperature ratio is set with an ad hoc prescription in post-processing \citep[e.g.][]{Shcherbakov12, Dexter12}; the most common prescription \citep{Moscibrodzka16} used to interpret EHT images sets the temperature ratio $R=T_{\rm i}/T_{\rm e}$ as a two-parameter function of the local ratio of the gas to magnetic pressure $\beta_{\rm gas}=\pgas/p_{\rm mag}$.
A lack of direct constraints on the electron temperature $T_{\rm e}$ from simulations is a limiting source of theoretical uncertainty in interpreting EHT observations.

\citet{Ressler15} introduced a new method to constrain $T_{\rm e}$ directly in black hole accretion simulations by extending the equations of GRMHD to include the thermodynamic evolution of separate electron and ion species.
For M87*, the moderately high radiative efficiency \citep{PaperVIII} means that radiative cooling is significant in determining $T_{\rm i}/T_{\rm e}$, so it is important to extend the two-temperature simulations further to include radiative feedback \citep{Sadowski17,Liska24}.
Two-temperature simulations of Sgr A* \citep{Ressler17, Chael18, Dexter20, Yoon20, Salas24, Moscibrodzka24} and M87* \citep{Ryan18, Chael19,Yao21} have successfully reproduced horizon-scale total intensity images consistent with post-processing models and EHT results.

While two-temperature, radiative GRMHD (2TGRRMHD) simulations have the potential to directly constrain the temperature of the electrons emitting the observed near-horizon radiation, these simulations have drawbacks when compared to the standard single-fluid GRMHD approach.
Because radiative coupling is usually handled implicitly \citep{Sadowski17,Liska22}, 2TGRRMHD simulations are significantly more computationally expensive than standard GRMHD evolution. Furthermore, adding radiation feedback breaks the scale-freedom of the ideal GRMHD equations; 2TGRRMHD simulations cannot be re-scaled freely in post-processing to match sources with different black hole masses and accretion rates. Finally, while the physics of electron cooling is well constrained, it is still uncertain exactly how electrons are heated and accelerated on microscopic scales around black holes.
2TGRRMHD parameterize this uncertainty in the choice of sub-grid electron heating function, $\delta_{\rm e}$, which determines how much of the dissipated energy heats electrons as a function of local plasma parameters. Two main families of models for $\delta_{\rm e}$ have been studied in 2TGRRMHD simulations, with electron heating originating from a turbulent cascade truncated by Landau damping \citep{Howes10,Kawazura19} or from sub-grid magnetic reconnection with or without a substantial guide field \citep{Rowan17,Rowan19}.

Because of the additional computational expense, single-source limitations, and uncertainty in the underlying electron heating mechanism, relatively few two-temperature GRMHD simulations of M87* have been conducted. In particular, there has not yet been an attempt to produce 2TGRRMHD versions of the simulation image ``libraries'' spanning different values of the black hole spin $a_*$ and electron temperature parameters that are the bedrock tool of analysis of polarized EHT observations with single-fluid GRMHD simulations \citep{PaperV,PaperVIII,SgrAPaperV,SgrAPaperVIII}.
In this paper, we begin a systematic series of investigations of a survey of 2TGRRMHD simulations of magnetically arrested models of M87*. We present a survey of 2TGRRMHD simulations of MAD accretion on a $M=6.5\times10^9M_\odot$ black hole with the accretion rate tuned to produce M87*'s 230 GHz observed flux density $F_{230}=0.5$ Jy \citep{PaperIV,EHT2024M87}. We compare the properties of our 2TGRRMHD simulations to fiducial single-fluid GRMHD simulations run with the same code and parameters, investigate the resulting distributions of electron and ion temperatures, and inspect the features of the resulting simulated 230 GHz images in total intensity and polarization across different values of black hole spin.
In this paper, we explore only one choice of the electron heating function $\delta_{\rm e}$, the \citet{Kawazura19} (\texttt{K19}) model of heating from plasma turbulence. We will compare the simulations presented here with analogous simulations conducted using the \citet{Rowan19} (\texttt{R19}) prescription for heating from magnetic reconnection in a future work.

The paper is organized as follows. In \autoref{sec:method} we review the equations and method of 2TGRRMHD evolution in the code \texttt{KORAL} \citep{Sadowski13}. In \autoref{sec:sims} we define the parameters of our simulation suite. In \autoref{sec:results} we investigate the results of our 2TGRRMHD simulations, comparing them across different values of the black hole spin and their non-radiative counterparts; we also investigate their resulting 230 GRMHD images and compare them to EHT observations of M87*. We discuss and conclude in \autoref{sec:discussion}.

\section{Two-Temperature, Radiative GRMHD Equations}
\label{sec:method}

\subsection{Units}
Except where otherwise noted, we use units where $G=c=1$. Length and time are measured relative to the gravitational length-scale $r_{\rm g}=GM/c^2$ and time-scale $t_{\rm g}=GM/c^3$, where $M$ is the black hole mass. The black hole accretion rate is measured relative to the Eddington accretion rate  $\dot{M}_\mathrm{Edd}=L_\mathrm{Edd}/\eta_{\rm Edd}c^2=4\pi \, GMm_{\rm p} / \eta_{\rm Edd} c  \sigma_{\rm T}$, where $m_{\rm p}$ is the proton mass, $\sigma_{\rm T}$ is the Thomson cross section, and we adopt an efficiency parameter $\eta_{\rm Edd}=0.1$. We frequently refer to the dimensionless black hole spin parameter $a_*=J/M^2$, where $J$ is the black hole's angular momentum.

For our simulations targeted at M87*, we adopt a black hole mass $M=6.5\times10^9 M_\odot$ and we assume the distance to M87* is $D=16.8$ Mpc \citep{PaperVI}.
For this choice of black hole mass, $r_{\rm g}=9.6\times10^{14}\,\mathrm{cm}=64\,\mathrm{AU}$, $t_{\rm g}=3.2\times10^4\,\mathrm{s}=8.9 \,\mathrm{hr}$, and $\dot{M}_{\rm Edd}=9.1\times10^{27}\,\mathrm{g \, s}^{-1}=144.3 \, M_{\rm \odot} \, \mathrm{yr}^{-1}$. The projected size of the gravitational radius is $\theta_{\rm g}=r_{\rm g}/D=3.82\,\mu$as.

\subsection{Equations}
Here we review the equations of two-temperature, radiative GRMHD used in the code \texttt{KORAL} \citep{Sadowski17}. The code evolves two fluids representing electrons ($\mathrm{e}$) and ions ($\mathrm{i}$) in a fully-ionized plasma.

The number densities of electrons and ions are related to the fluid rest-mass density $\rho$ by $n_{\rm e,i}=\rho/\left(\mu_{\rm e,i}m_{\rm p}\right)$, where $m_{\rm p}$ is the proton mass and $\mu_{\rm e,i}$ is the electron (ion) mean molecular mass. In this work, we assume the gas is fully ionized hydrogen, so $\mu_{\rm e}=\mu_{\rm i}=1$.
The electron and ion pressures $p_{\rm e,i}$ are related to their respective temperatures  $T_{\rm e,i}$ by the ideal gas law, $p_{\rm e,i} = n_{\rm e,i} k_{\rm B} T_{\rm e,i}$. The electron/ion internal energies are then given by an equation of state:
\begin{align}
u_{\rm e,i} &= \frac{p_{\rm e,i}}{\Gamma_{\rm e,i}(\Theta_{\rm e,i})-1},
\end{align}
where the dimensionless temperature for electrons and ions is $\Theta_{\rm e,i}=k_{\rm B}T_{\rm e,i}/m_{\rm e,i}c^2$, where $m_{\rm e}$ is the electron mass and $m_{\rm i}=\mu_{\rm i}m_{\rm p}$ is the mean ion mass.  We assume both species are in a relativistic Maxwell-J\"uttner distribution;
$\Gamma(\Theta)$ is a temperature-dependent adiabatic index that transitions from $\Gamma=5/3$ for a non-relativistic fluid, $\Theta\ll1$, to $\Gamma=4/3$ for a relativistic fluid, $\Theta\gg1$. The form of $\Gamma(\Theta)$ we use is given in \citet{Sadowski17}, Equation A14.

We assume the electron and ion fluids combine to form an effective single fluid with total pressure $\pgas=p_{\rm e}+p_{\rm i}$ and energy density $\ugas = u_{\rm e}+u_{\rm i}$. The effective combined gas temperature $T_{\rm gas}$ and adiabatic index $\adi$ are then
\begin{align}
\label{eq::gammaeff}
T_\mathrm{gas} &= \mu_{\rm gas}\left(\frac{T_\mathrm{i}}{\mu_{\rm i}} + \frac{T_\mathrm{e}}{\mu_{\rm e}}\right), \\
\adi &= 1 + \frac{\left(\Gamma_\mathrm{i} -1\right)\left(\Gamma_\mathrm{e} -1\right)\left(T_\mathrm{i}/T_\mathrm{e} + \mu_{\rm i}/\mu_{\rm e}\right)}{\left(T_\mathrm{i}/T_\mathrm{e}\right)\left(\Gamma_\mathrm{e}-1\right) + \left(\mu_{\rm i}/\mu_{\rm e}\right)\left(\Gamma_\mathrm{i} -1\right)},
\end{align}
where the effective mean molecular mass for the combined two-fluid plasma is $\mu^{-1}_{\rm gas} = \mu_{\rm e}^{-1}+\mu_{\rm i}^{-1}$; $\mu_{\rm gas}=1/2$ in our simulations assuming pure Hydrogen.

The combined electron-ion fluid and electromagnetic field together make up the MHD stress-energy tensor $T^\mu_{\;\;\nu}$;
\begin{align}
 \label{eq::tmunu}
 T^\mu_{\;\;\nu} &= \left(\rho + \ugas + \pgas + b^2\right)u^\mu u_\nu + \left(\pgas +
\frac{1}{2}b^2\right)\delta^\mu_{\;\;\;\nu} - b^\mu b_\nu.
\end{align}
The fluid four-velocity is $u^\mu$ and the fluid-frame magnetic field vector is $b^\mu=u_\nu \star F^{\nu\mu}$, where $\star F^{\nu\mu}$ is the dual of the electromagnetic field Faraday tensor $F^{\nu\mu}$. We take the ideal MHD limit, where the fluid is assumed to be infinitely conductive so that the fluid-frame electric field vanishes, $e^\mu=u_\nu F^{\mu\nu}=0$. In the ideal MHD limit, $b^\mu$ can be computed from $u^\mu$ and the ``lab-frame'' magnetic field $B^i=\star F^{i0}$ in the standard way \citep[e.g.][]{Gammie03}.

In addition to the plasma and magnetic field, we also evolve a radiation field coupled to the plasma using the M1 approximation, where we assume there always exists a timelike frame  $u^\mu_{\rm R}$ where the radiation stress-energy $R^\mu_{\;\;\nu}$ is isotropic. The radiation field stress-energy tensor in the M1 approximation is
\begin{align}
\label{eq::rmunu}
 R^\mu_{\;\;\nu} &= \frac{4}{3}\bar{E}u^\mu_{\rm R} u^{\vphantom{\mu}}_{\rm R\,\nu} +
\frac{1}{3}\bar{E}\delta^\mu_{\;\;\nu},
\end{align}
where $\bar{E}$ is the radiation energy density in the frame co-moving with $u^\mu_{\rm R}$. To account for frequency-dependence in emission and absorption processes to first order,  \texttt{KORAL} also tracks the photon number density $\bar{n}_{\rm R}$, which provides the mean photon frequency in the fluid frame  $\hat{E}/h\hat{n}_{\rm R}$, or equivalently, the effective radiation blackbody temperature $T_{\rm R}$ \citep{Sadowski15}.\footnote{After transforming both the radiation energy density and photon number density to the fluid frame (throughout, hats indicate fluid frame quantities and bars indicate radiation-frame quantities),
$\hat{n}_{\rm R} = -\bar{n}_{\rm R}u^\mu_{\rm R} u_\mu$,  $\hat{E}=R^{\mu\nu}u_\mu u_\nu$, the radiation colour temperature is $T_{\rm R}=\hat{E}/2.7012k_{\rm B}\hat{n}_{\rm R}$.}

The coupled evolution equations for the two-temperature fluid, the electromagnetic field, and the radiation field are
\begin{align}
\label{eq::GRRMHD}
\nabla_\mu(\rho u^\mu) & = 0, \\
\nabla_\mu \left(T^{\mu\nu}\right) &= G_\nu, \\
\nabla_\mu \left(R^{\mu\nu}\right) &= -G_\nu, \\
\label{eq::GRRMHD4}
\nabla_\mu\left(\star F^{\mu\nu}\right) &=0,
\end{align}
where $G^\nu$ is the radiation four-flux, obtained from the emissivities and opacities for the radiation processes considered using the formulae in \citep{Sadowski17}. In our simulations, we consider radiation from synchrotron, bremsstrahlung, Thomson and Compton scattering processes, though synchrotron emission is the main contribution to $G^\nu$ \citep{Salas24}.
The evolution equation for the photon number density is
\begin{equation}
\label{eq::photev}
\nabla_\mu \left(\bar{n}^{\vphantom{\mu}}_{\rm R} u^\mu_{R}\right) = \dot{\bar{n}}_{\rm R} ,
\end{equation}
where $\dot{\bar{n}}_{\rm R}$ is the photon production rate in the radiation frame \citep[see][]{Sadowski15,Sadowski17}.

To evolve the individual energy densities of the electron and ion fluids, we use the first law of thermodynamics for each with appropriate source terms;
\begin{align}
\label{eq::ent_ev}
T_\mathrm{e}\nabla_\mu\left(n_{\rm e} s_\mathrm{e} u^\mu\right) &= \delta_\mathrm{e} q^{\rm v} +
q^{\rm C} - \hat{G}^0,  \\
\label{eq::ent_ev2}
T_\mathrm{i}\nabla_\mu\left(n_{\rm i} s_\mathrm{i} u^\mu\right) &= (1-\delta_\mathrm{e}) q^{\rm v} - q^{\rm C}.
\end{align}
In \autoref{eq::ent_ev} and \autoref{eq::ent_ev2},
$s_{\rm e,i}$ is the temperature-dependent entropy per particle of the electron and ion species (see Equation A11 of \citealt{Sadowski17} for the form of $s(n,\Theta)$ that we use), $q^{\rm v}$ is the viscous heating rate, $q^{\rm C}$ is the Coulomb coupling rate \citep{Stepney83}, and $\hat{G}^0$ is the radiation cooling rate transformed to the fluid frame. The parameter $\delta_{\rm e}$ controls the fraction of the total dissipated energy that goes into electrons.

As in \citep{Sadowski17,Chael19,Salas24}, we compute the viscous heating rate $q^{\rm v}$ numerically by first evolving \autoref{eq::ent_ev} and \autoref{eq::ent_ev2} without source terms and then comparing the sum of the adiabatically evolved electron and ion energy densities to the separately-evolved total fluid energy $\ugas$. To determine the electron heating fraction $\delta_{\rm e}$, we adopt the \citet{Kawazura19} (hereafter \texttt{K19}) prescription from gyrokinetic simulations of plasma turbulence.
The \citet{Kawazura19} heating prescription gives $\delta_{\rm e}$ as a function of the local electron-to-ion temperature ratio and plasma-beta:
\begin{align}
\label{eq::kawazura}
\frac{Q_{\rm i}}{Q_{\rm e}} &= \frac{35}{\left(\beta_{\rm i}/15\right)^{-1.4}e^{-0.1T_{\rm e}/T_{\rm i}}}, \\
\delta_{\rm e} &= \frac{1}{1+Q_{\rm i}/Q_{\rm e}},
\end{align}
where the plasma-beta parameter for the ions $\beta_{\rm i}$ is  the ratio of the ion to magnetic pressure in the fluid frame:
\begin{equation}
\label{eq:betai}
\beta_{\rm i} = \frac{p_{\rm i}}{p_{\rm mag}} = \frac{2n_{\rm i} k_{\rm B} T_{\rm i}}{b^2}.
\end{equation}
Note that $\beta_{\rm i}$ defined in \autoref{eq:betai} is the plasma-beta parameter defined relative to the ion temperature; we also make use of a similarly-defined electron beta $\beta_{\rm e}=p_{\rm e}/p_{\rm mag}$ and the standard plasma-beta for the combined electron-ion gas, $\beta_{\rm gas}=\pgas/p_{\rm mag} = \beta_{\rm e}+\beta_{\rm i}$.
We also often make use of the plasma magnetization parameter, which represents the ratio of the magnetic energy density to the ion rest mass energy density:
\begin{equation}
\sigma_{\rm i} = \frac{b^2}{n_{\rm i} m_{\rm i} c^2}.
\end{equation}

\section{Simulations}
\label{sec:sims}
We ran a total of 11 radiative two-temperature GRMHD simulations of M87* and 11 corresponding benchmark ideal single-fluid simulations in the code \texttt{KORAL} \citep{Sadowski13,Sadowski17}. All of our simulations were conducted in the magnetically arrested (MAD) state, and we span a range of black hole spins $a_*\in\{\pm0.9,\pm0.7,\pm0.5,\pm.0.3,\pm0.1,0\}$. We list the simulations considered here in \autoref{tab:model_summary}.

\begin{table*}
    \centering
    \begin{tabular}{l|cccc|ccc|ccccc}
        \toprule
        Model  & $a_*$ & $B$-field & Radiation & Electron Heating &  $N_r \times N_\theta \times N_\phi$ & $r_{\rm min}/r_{\rm g}$ & $r_{\rm max}/r_{\rm g}$& $\langle \dot{M}\rangle / 10^{-7}\dot{M}_{\rm Edd}$ & $\langle \phi_{\rm BH} \rangle$ & $\langle \eta \rangle$ & $\langle s \rangle$ & $\langle \epsilon_{\rm rad} \rangle /10^{-2}$ \\
        \midrule
\texttt{ap9} & 0.9 & MAD & No & - & $288\times192\times144$ & 1.18 & $10^4$ & 9.8 & 46.7 & 1.2 & -8.1 & - \\
\texttt{ap7} & 0.7 & MAD & No & - & $288\times192\times144$ & 1.41 & $10^4$ & 13.9 & 60.5 & 0.9 & -9.4 & - \\
\texttt{ap5} & 0.5 & MAD & No & - & $288\times192\times144$ & 1.54 & $10^4$ & 20.4 & 59.0 & 0.4 & -6.0 & - \\
\texttt{ap3} & 0.3 & MAD & No & - & $288\times192\times144$ & 1.61 & $10^4$ & 24.6 & 60.7 & 0.2 & -3.7 & - \\
\texttt{ap1} & 0.1 & MAD & No & - & $288\times192\times144$ & 1.65 & $10^4$ & 31.7 & 53.4 & 0.1 & -0.6 & - \\
\texttt{ap0} & 0.0 & MAD & No & - & $288\times192\times144$ & 1.65 & $10^4$ & 30.8 & 57.1 & 0.0 & 0.6 & - \\
\texttt{am1} & -0.1 & MAD & No & - & $288\times192\times144$ & 1.65 & $10^4$ & 28.8 & 52.2 & 0.1 & 1.9 & - \\
\texttt{am3} & -0.3 & MAD & No & - & $288\times192\times144$ & 1.61 & $10^4$ & 26.3 & 45.8 & 0.1 & 3.6 & - \\
\texttt{am5} & -0.5 & MAD & No & - & $288\times192\times144$ & 1.54 & $10^4$ & 29.1 & 36.8 & 0.1 & 4.4 & - \\
\texttt{am7} & -0.7 & MAD & No & - & $288\times192\times144$ & 1.41 & $10^4$ & 15.9 & 28.8 & 0.2 & 4.7 & - \\
\texttt{am9} & -0.9 & MAD & No & - & $288\times192\times144$ & 1.18 & $10^4$ & 16.3 & 23.4 & 0.3 & 5.1 & - \\
\midrule
\texttt{ap9\_radk} & 0.9 & MAD & Yes & \texttt{K19} & $288\times192\times144$ & 1.18 & $10^4$ & 5.7 & 43.6 & 1.2 & -6.8 & 32.9 \\
\texttt{ap7\_radk} & 0.7 & MAD & Yes & \texttt{K19} & $288\times192\times144$ & 1.41 & $10^4$ & 8.1 & 57.8 & 0.9 & -8.3 & 24.3 \\
\texttt{ap5\_radk} & 0.5 & MAD & Yes & \texttt{K19} & $288\times192\times144$ & 1.54 & $10^4$ & 8.9 & 57.5 & 0.4 & -5.7 & 12.6 \\
\texttt{ap3\_radk} & 0.3 & MAD & Yes & \texttt{K19} & $288\times192\times144$ & 1.61 & $10^4$ & 10.0 & 63.1 & 0.2 & -4.1 & 10.0 \\
\texttt{ap1\_radk} & 0.1 & MAD & Yes & \texttt{K19} & $288\times192\times144$ & 1.65 & $10^4$ & 12.5 & 57.1 & 0.1 & -0.9 & 4.6 \\
\texttt{ap0\_radk} & 0.0 & MAD & Yes & \texttt{K19} & $288\times192\times144$ & 1.65 & $10^4$ & 11.3 & 54.2 & 0.1 & 0.5 & 3.6 \\
\texttt{am1\_radk} & -0.1 & MAD & Yes & \texttt{K19} & $288\times192\times144$ & 1.65 & $10^4$ & 14.8 & 50.2 & 0.1 & 1.6 & 4.4 \\
\texttt{am3\_radk} & -0.3 & MAD & Yes & \texttt{K19} & $288\times192\times144$ & 1.61 & $10^4$ & 15.5 & 45.1 & 0.1 & 3.1 & 4.3 \\
\texttt{am5\_radk} & -0.5 & MAD & Yes & \texttt{K19} & $288\times192\times144$ & 1.54 & $10^4$ & 13.3 & 37.5 & 0.1 & 3.7 & 5.4 \\
\texttt{am7\_radk} & -0.7 & MAD & Yes & \texttt{K19} & $288\times192\times144$ & 1.41 & $10^4$ & 9.2 & 36.7 & 0.2 & 4.8 & 9.3 \\
\texttt{am9\_radk} & -0.9 & MAD & Yes & \texttt{K19} & $288\times192\times144$ & 1.18 & $10^4$ & 13.3 & 27.2 & 0.3 & 4.6 & 11.4 \\
\bottomrule
    \end{tabular}
    \caption{Summary of 3D MAD simulations presented in this work.
    For each simulation we report the spin $a_*$, the magnetic field configuration (all simulations are in the MAD state), whether the simulation includes radiation feedback, and the electron heating function (we use only the \citet{Kawazura19} function \texttt{K19} in this work). We also report the simulation resolution and the inner and outermost simulation radii $r_{\rm min}, r_{\rm max}$. Over the final $[15000,20000]\,t_{\rm g}$ interval of each simulation we compute the mean
    mass accretion rate through the horizon $\dot{M}$ (\autoref{eq:mdot}) in Eddington units, the normalized horizon magnetic flux $\phi_{\rm BH}$ (\autoref{eq:madpar}), the outflow efficiency $\eta$ (\autoref{eq:jeteff}), the spindown parameter $s$ (\autoref{eq:spindown}), and for the radiative simulations, the radiative efficiency $\epsilon_{\rm rad}$ (\autoref{eq:radeff}).
    The values of $\dot{M}/\dot{M}_{\rm Edd}$ for the non-radiative models reported here correspond to a density scaling used to match a median 230 GHz flux density $F_{230}=0.5$ Jy for each simulation using a $R_{\rm low}=10, R_{\rm high}=40$ \citet{Moscibrodzka16} electron temperature model.
    }
    \label{tab:model_summary}
\end{table*}

\subsection{Numerical Implementation}
We concentrate resolution in the jet region close to the polar axis and the disc region near the equator by transforming standard Kerr-Schild spatial coordinates $(r,\theta,\phi)$ to the ``\texttt{jetcoords}'' coordinate system $(x_1,x_2,x_3)$ introduced in \citet{Ressler17} and used in \citet{Narayan22}. Each simulation has a resolution of $288,192,144$ in the uniform grid of code coordinates $(x_1,x_2,x_3)$. The radial coordinate $x_1$ is logarithmically spaced from $r_{\rm min}=0.825r_+$ to $r_{\rm max}=10^4\,r_{\rm g}$, where $r_+$ is the radius of the event horizon. The simulation grid is uniform in $x_3=\phi$. The polar angle $\theta$ is a function of both $x_1$ and $x_2$; we use the same parameters for this transformation as \citet{Narayan22}, except that the minimum value of the $x_2$ coordinate here is $x_{2,\mathrm{min}}=10^{-3}$.

We initialized the single-fluid simulations with a \citet{FishMonc} equilibrium torus with a constant inner radius $r_{\rm in}=20\,r_{\rm g}$. Because the outer edge of the \citet{FishMonc} torus is highly sensitive to the combination of black hole spin, inner edge $r_{\rm in}$, and radius of maximum pressure $r_{p,\mathrm{max}},$ we follow \citet{Narayan22} and choose a
spin-dependent pressure maximum radius in the range $r_{p,\mathrm{max}}\sim40-42\,r_{\rm g}$ such that our initial tori all have outer edges inside the simulation domain. To evolve the system to the MAD state, we initialized the torus with a coherent loop of poloidal magnetic field centred exactly as in \citet{Narayan22}.  The magnetic field strength in the disc was normalized so that the ratio of the maximum gas pressure to maximum magnetic pressure in the torus $p_{\rm gas,max}/p_{\rm mag,max}=100$.

Like most GRMHD codes \citep[e.g.][]{Gammie03}, \texttt{KORAL} solves the GRMHD equations in a finite-volume scheme. \texttt{KORAL} reconstructs primitive quantities at cell walls using the second-order piecewise parabolic method (PPM) and uses the local Lax-Friedrichs method to evaluate fluxes. The coupling $G^\nu$ between the fluid and radiation is handled implicitly, and the time evolution is performed with a second-order Implicit-Explicit (IMEX) scheme.
The magnetic field evolution is constrained to conserve the initially zero divergence with the constrained transport algorithm of \citet{Toth00}. Our simulations use outflowing boundary conditions at the inner boundary $r_{\rm min}$ inside the event horizon and at $r_{\rm max}$. The azimuthal boundary is periodic. We use reflecting boundary conditions at the polar axis, with a modification to the $x_2$-component of the plasma (or radiation-frame) velocity in the inner two cells such that this component goes to zero at the pole.

We first evolved the 11 single-fluid GRMHD simulations from Kerr-Schild time $t=0$ to $t=10^4\,t_{\rm g}$. During this initial period, gas in the torus begins to accrete and the magnetic field builds up on the black hole, eventually saturating at a value of the dimensionless magnetic flux (or ``MAD parameter'') $\phi_{\rm BH}\approx50$, where $\phi$ is given by \autoref{eq:madpar}. At this point, we initialized the eleven two-temperature, radiative simulations from the already-accreting single-fluid GRMHD data. We ran both the parent single-fluid simulations and the two-temperature simulations from this point to a maximum time $t=2\times10^4\,t_{\rm g}$.

When re-starting the single fluid GRMHD simulations to evolve the 2TGRRMHD equations, we initialize the radiation energy density $\bar{E}=10^{-4}\ugas$, and the electron energy density $u_{\rm e}=0.05\ugas$, (so $u_{\rm i}=0.95\ugas$).
Single-fluid GRMHD simulations are scale-free, but radiative simulations require a fixed choice of physical energy density scale.
We first scale the density and energy densities so that the average black hole accretion rate in the range $t\in [9000,10000]\,t_{\rm g}$  is $\dot{M}=10^{-6}\dot{M}_{\rm Edd}$, an approximate value for M87* MAD models \citep{PaperV}. We then run the simulation for an additional $500 \,t_{\rm g}$ so the radiation and two-temperature effects in the innermost accretion flow can reach an approximate steady-state. We then re-scale the simulation density and energy densities a second time so that the median 230 GHz flux density from synchrotron emission in the last $250\,t_{\rm g}$ is $F_{230}=0.5$ Jy \citet{PaperIV}. We finally ran the radiative simulations from this point forward without additional density rescalings during the simulation runtime to the maximum time of $t=2\times10^4\,t_{\rm g}$, but when producing 230 GHz images for comparison to EHT observations we apply a small post-processing density rescaling to account for secular drifts in $F_{230}$ (see \autoref{sec:images1}).
We save snapshots of each simulation every $10\,t_{\rm g}$.

In our radiative, two-temperature simulations the adiabatic index is adjusted self-consistently in each cell at each time step to reflect the local ion and electron temperatures through \autoref{eq::gammaeff}.
In our single-fluid benchmark GRMHD simulations, we use a fixed adiabatic index $\adi=13/9$ throughout, corresponding to $T_{\rm e}=T_{\rm i}$, $\Gamma_{\rm e}=4/3$, $\Gamma_{\rm i}=5/3$.  All of our radiative simulations use the \texttt{K19}
electron heating function \autoref{eq::kawazura}.

While we have recently added the option for hybrid GRMHD and Force-Free evolution for magnetized accretion flows in \texttt{KORAL} \citep{Chael24}, for this simulation suite we use the standard approach of applying density floors in our simulation to ensure numerical stability in highly-magnetized regions. \texttt{KORAL} injects rest mass in the zero-angular observer (ZAMO) frame to enforce a maximum magnetization $\sigma_{\rm i}\leq\sigma_{\rm max}=100$. To ensure numerical stability, \texttt{KORAL} also applies other floors and ceilings on the plasma (and radiation) energy densities, the plasma (and radiation-frame) Lorentz factor, and on the electron and ion temperatures \citep{Sadowski17,Porth19}.

\subsection{230 GHz Images}
\label{sec:images1}
\begin{table}
    \centering
    \begin{tabular}{l|l}
        \toprule
        Model & $f_{230}$ \\
        \midrule
        \texttt{ap9\_radk} & 1.27 \\
        \texttt{ap7\_radk} & 0.99 \\
        \texttt{ap5\_radk} & 1.23 \\
        \texttt{ap3\_radk} & 1.23 \\
        \texttt{ap1\_radk} & 1.04 \\
        \texttt{a0\_radk}  & 1.08 \\
        \texttt{am1\_radk} & 0.91 \\
        \texttt{am3\_radk} & 0.98 \\
        \texttt{am5\_radk} & 1.02 \\
        \texttt{am7\_radk} & 0.93 \\
        \texttt{am9\_radk} & 0.94 \\
        \bottomrule
    \end{tabular}
    \caption{Secondary density rescaling factors for radiative simulations.
    }
    \label{tab:rescaling}
\end{table}

To simulate EHT observations of M87*, we produce 230 GHZ images from the eleven radiative simulations using the GR radiative transfer code \texttt{ipole} \citep{ipole}. For each simulation we produce 500 snapshot images in full polarization covering the time range $15000-20000\,t_{\rm gc}$. Following \citet{PaperV}, we fix the observer inclination of the prograde simulations (including $a_*=0$) to $\theta_{\rm o}=163\deg$ and the viewing inclination of the retrograde simulations to $\theta_{\rm o}=17\deg$ \citep{Mertens2016}. We orient the images so the black hole spin faces East (a position angle $\varphi=90\deg$ counter-clockwise from vertical). We exclude emission from highly magnetized regions with $\sigma_{\rm i}\geq\sigma_{\rm cut}=25$.
The field of view of our images is $150\,\mu$as ($\approx\,39\theta_{\rm g}$), and we use a pixel size of $0.3\,\mu$as.

In its 2017 and 2018 observations of M87*, the EHT constrained the 230 GHz compact flux density on horizon scales to be $F_{230}\approx0.5$ Jy \citep{PaperIV,EHT2024M87}. As a result, GRMHD simulations of M87* are typically scaled so that the median 230 GHz flux density over some time period matches the observed $0.5$ Jy value. This presents no problem for scale-free pure GRMHD simulations, but introducing radiation fixes a physical density scale which in principle should not be adjusted at all in post-processing. Our radiative simulations were rescaled to achieve $F_{230}\approx0.5$ Jy at $t=10500\,t_{\rm g}$, but the simulations do not maintain exactly the same median 230 GHz flux density over a wide time window due to secular drifts in the accretion rate and other quantities. In order to provide a direct comparison across the simulations that is consistent with standard practice in the field, the results shown here apply a second small density rescaling to the simulations in post-processing, chosen such that the median $F_{230}=0.5$ Jy in the window $15000-20000\,t_{\rm g}$. The applied post-processing rescaling factor $f_{\rm 230}$ for each simulation is reported in \autoref{tab:rescaling}. Most of the simulations have their density (and internal energy density, magnetic energy density, and radiation energy density) scaled by a factor $<10\%$, with a maximum adjustment of 27\% for simulation \texttt{ap9\_radk}. None of the results presented below change qualitatively when comparing results with and without this post-processing density rescaling.

We also generate images from our non-radiative GRMHD simulations for comparison with the radiative simulations. To set the electron temperature in the images from the non-radiative simulations, we use the \citet{Moscibrodzka16} $R_{\rm high},R_{\rm low}$ prescription that adjusts the ion-to-electron temperature ratio $R=T_{\rm i}/T_{\rm e}$ based on the local $\beta_{\rm gas}$. We explore a variety of $R_{\rm low},R_{\rm high}$ combinations in the range $R_{\rm low}\in\{1,10\},R_{\rm high}\in\{1,10,20,40,80,160\}.$

\section{Results}
\label{sec:results}

All results presented here are taken from the last $5000\,t_{\rm g}$ of evolution from $t=15000\,t_{\rm g}$ to $t=20000\,t_{\rm g}$.

\subsection{Average Fluxes}
\begin{figure*}
\centering
\includegraphics[width=\textwidth]{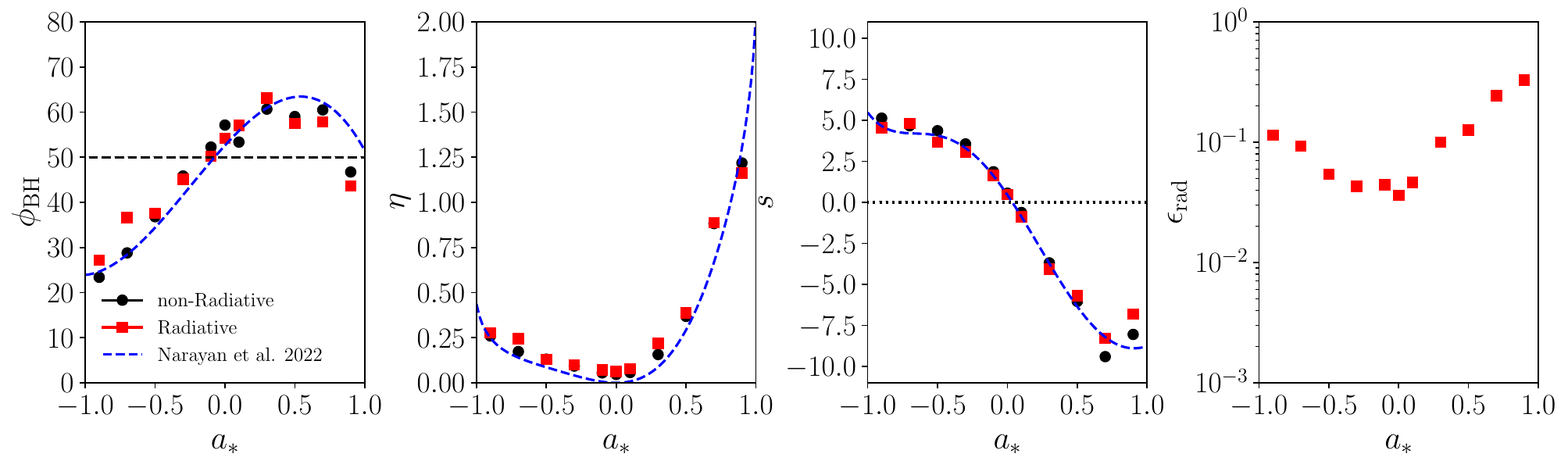}
\caption{Dimensionless simulation fluxes. The first panel shows the average value of the dimensionless magnetic flux, or ``MAD parameter'' $\phi_{\rm BH}$ (\autoref{eq:madpar}) for all simulations reported in this work as a function of black hole spin $a_*$. The second panel shows the outflow efficiency factor $\eta$ (\autoref{eq:jeteff}) for all simulations, and the third panel plots the spindown parameter $s$ (\autoref{eq:spindown}). The final panel plots the bolometric radiative efficiency $\epsilon_{\rm rad}$ (\autoref{eq:radeff}) for the radiative simulations.
In all panels, results are averaged over the window $t\in\left[15000,20000\right]t_{\rm g}$, and black circles show results from the non-radiative simulations while red squares show results from the radiative, two-temperature simulations. In the first three panels, we plot fitting functions for $\phi_{\rm BH}$, $\eta$, and $s$ derived in \citet{Narayan22} in dashed blue lines; these correspond to Equations 9, 10, and 15 of \citet{Narayan22}, respectively.
}
\label{fig:fluxes}
\end{figure*}

We first compare the average fluxes of mass, magnetic field, energy, and angular momentum for the radiative simulations with their non-radiative counterparts.
At each saved snapshot in the range $t\in[15000,20000]t_{\rm g}$ we calculate the black hole accretion rate $\dot{M}$ and  magnetic flux through the horizon $\Phi_{\rm BH}$ using the following integrals over the horizon at $r=r_+$;
\begin{align}
\label{eq:mdot}
\dot{M} &= \int_\theta \int_\phi \left(-\rho u^r\right)\sqrt{-g}\;\ed\phi \ed\theta, \\
\Phi_{\rm BH} &= \int_\theta\int_\phi\left|B^r\right|\sqrt{-g}\;\ed\phi \ed\theta,
\end{align}
where $g$ is the metric determinant, and we first convert $u^\mu$ and $B^i$ from code coordinates to standard Kerr-Schild coordinates.
The dimensionless magnetic flux on the horizon, or ``MAD parameter'' is \citep{Tchekhovskoy11}:\footnote{Note that we follow \citet{Narayan22} in defining the MAD parameter in Gaussian units by including a factor of $\sqrt{4\pi}$.}
\begin{equation}
\label{eq:madpar}
\phi_{\rm BH} = \frac{\sqrt{4\pi}}{2\sqrt{\dot{M}}} \Phi_{\rm BH}.
\end{equation}
For accretion discs in the MAD state, the magnetic flux saturates at $\phi_{\rm BH}\approx50$ \citep{Tchekhovskoy11,McKinney12}, though the precise value is spin- and initial-condition-dependent \citep{Tchekhovskoy12,Narayan22}.

In \autoref{tab:model_summary}, we report the mean values of $\phi_{\rm BH}$ for each simulation over the range $t\in[15000,20000]t_{\rm g}$, and we plot the mean $\phi_{\rm BH}$ as a function of $a_*$ in the first panel of \autoref{fig:fluxes}. We find that the radiative and non-radiative simulations show a similar trend of $\phi_{\rm BH}$ with black hole spin; retrograde simulations have lower $\phi_{\rm BH}$ values which increase as $a_*\rightarrow0$; $\phi_{\rm BH}$ reaches a maximum of $\phi_{\rm BH}\approx60$ around $a_*\approx0.3-0.5$ before declining again at large positive spin values. This trend is consistent with what is observed in \citet{Narayan22} (from much longer ideal GRMHD simulations run from the same initial conditions with the same \texttt{KORAL} code) and \citet{Tchekhovskoy12} (from simulations run with a different code). We show the fitting function for $\phi_{\rm BH}(a_*)$ derived in \citet{Narayan22} in the left panel of \autoref{fig:fluxes} in the dashed blue line. Our results for both the radiative and non-radiative simulations are approximately consistent with the unadjusted \citet{Narayan22} fitting function; the addition of radiation to our M87* simulations does not significantly affect the average magnetic flux accumulated on the black hole.

We compute the radial angular momentum flux $\dot{J}$ and the radial energy flux $\dot{E}$ at radius $r=5\,r_{\rm g}$ in order to avoid effects from numerical floors close to the horizon, following \citet{Narayan22}. These are:
\begin{align}
\dot{J} &= \int_\theta\int_\phi -T^r_\phi\sqrt{-g}\;\ed\phi \ed\theta, \\
\dot{E} &= \int_\theta\int_\phi T^r_t\sqrt{-g}\;\ed\phi \ed\theta.
\end{align}
The ratio $\eta$ of the outflowing energy flux to the accretion rate is
\begin{equation}
\label{eq:jeteff}
\eta = \frac{\dot{M}-\dot{E}}{\dot{M}}.
\end{equation}
The parameter $\eta$ represents the efficiency of the energy output by the black hole-accretion system. In spinning MAD simulations, nearly all of the outgoing power is in outward electromagnetic Poynting flux in the jet driven by the \citet{BZ} process \citep{McKinney04,Tchekhovskoy11,McKinney12}; we thus sometimes refer to $\eta$ as a ``jet efficiency,'' though \autoref{eq:jeteff} makes no distinction between the outflowing power in the jet and slow-moving wind components \citep{Narayan22}.

The black hole spindown (or spinup) parameter $s$ represents the rate of change $a_*$ relative to the black hole accretion rate \citep{Shapiro05,Narayan22}:
\begin{equation}
\label{eq:spindown}
s = \frac{da_*}{dt}\frac{M}{\dot{M}}=\frac{\dot{J}-2a_*\dot{E}}{\dot{M}}.
\end{equation}
A value of $s$ that is the same sign as the black hole spin indicates that the accretion disc angular momentum is spinning the black hole up; conversely, if $s$ is the opposite sign to $a_*$ either the counter-rotating accretion disc or the jet outflow is spinning the black hole down.

In \autoref{tab:model_summary}, we report the mean values of $\eta$ and $s$ for each simulation over the range $t\in[15000,20000]t_{\rm g}$. We plot $\eta$ and $s$ vs black hole spin in the second and third panels of \autoref{fig:fluxes}. We find that the trend of $\eta$ and $s$ with spin is virtually unchanged by the addition of radiation and two-temperature physics. For the most rapidly spinning black holes with prograde accretion discs, we find $\eta>100\%$, indicating that the energy flowing out in the BZ jet exceeds the rate of rest-mass energy infall through the horizon; for less rapidly spinning prograde discs and for all retrograde MAD disc-jet systems, the jet efficiency $\eta<100\%$. We confirm that the sixth-order BZ result for $\eta(\phi)$ from \citet{Tchekhovskoy10} fits our measured efficiencies for both the radiative and non-radiative simulations well. The dashed blue line in the second panel of \autoref{fig:fluxes} is the same as in Figure 4 of \citet{Narayan22}, namely
\begin{equation}
\eta(\phi_{\rm BH}) = \frac{1}{8\pi}\phi^2_{\rm BH}\left[\Omega_{\rm H}^2 + 1.38 \Omega_{\rm H}^4 - 0.2\Omega_{\rm H}^6\right],
\end{equation}
where $\Omega_{\rm H}=a_*/2r_+$ is the angular frequency of the black hole horizon and we use the same fitting function for $\phi_{\rm BH}(a_*)$ as \citet{Narayan22}. The agreement between the simulation results for $\eta(a_*)$ and the BZ prediction given the measured magnetic flux $\phi_{\rm BH}$ suggests that most of the outflowing energy is extracted from the BH as Poynting flux in the \citet{BZ} process.

Our results for the spindown parameter $s(a_*)$ in the third panel of \autoref{fig:fluxes} also show few differences between the radiative and non-radiative simulations, and both match the previous fitting function derived in \citet{Narayan22} (blue dashed line). As in that previous work, we find that for all simulations with non-zero spin the spindown parameter $s$ has the opposite sign to $a_*$, suggesting that the black hole jet is extracting angular momentum from the black hole more efficiently than it can be supplied by prograde accretion. Thus, jets from magnetically arrested black holes tend to drive the spin to zero if the black hole remains in the magnetically arrested state \citep{Narayan22, Ricarte23}.

For the radiative simulations, we also compute the bolometric luminosity $\dot{L}$:
\begin{align}
L_{\rm bol} &= \int_\theta\int_\phi -R^r_t\sqrt{-g}\;\ed\phi \ed\theta.
\end{align}
We compute the bolometric luminosity $\dot{L}$ at a larger radius $r=100\,r_{\rm g}$ than we use in computing $\dot{E}$ and $\dot{J}$ in order to account for radiation produced throughout the inner disk and jet regions.
The radiative efficiency $\epsilon_{\rm rad}$ is then the ratio of the bolometric luminosity computed at $r=100\,r_{\rm g}$ to the accretion rate at $r=r_+$:
\begin{equation}
\label{eq:radeff}
\epsilon_{\rm rad} = \frac{L_{\rm bol}}{\dot{M}}.
\end{equation}
In \autoref{tab:model_summary}, we report average values of $\epsilon_{\rm rad}$, and we plot $\epsilon_{\rm rad}$ against black hole spin in the rightmost panel of \autoref{fig:fluxes}. The radiative efficiencies in all of our M87* MAD simulations range from 3-35\%, in the range consistent with the results from the survey of non-radiative MAD simulations in \citet{PaperVIII}. We find that the radiative efficiency increases with both prograde and retrograde spin, though the rapidly spinning prograde models have higher values of $\epsilon_{\rm rad}$ than the rapidly spinning retrograde models.

\subsection{Disc Profiles}

\begin{figure*}
\centering
\includegraphics[width=\textwidth]{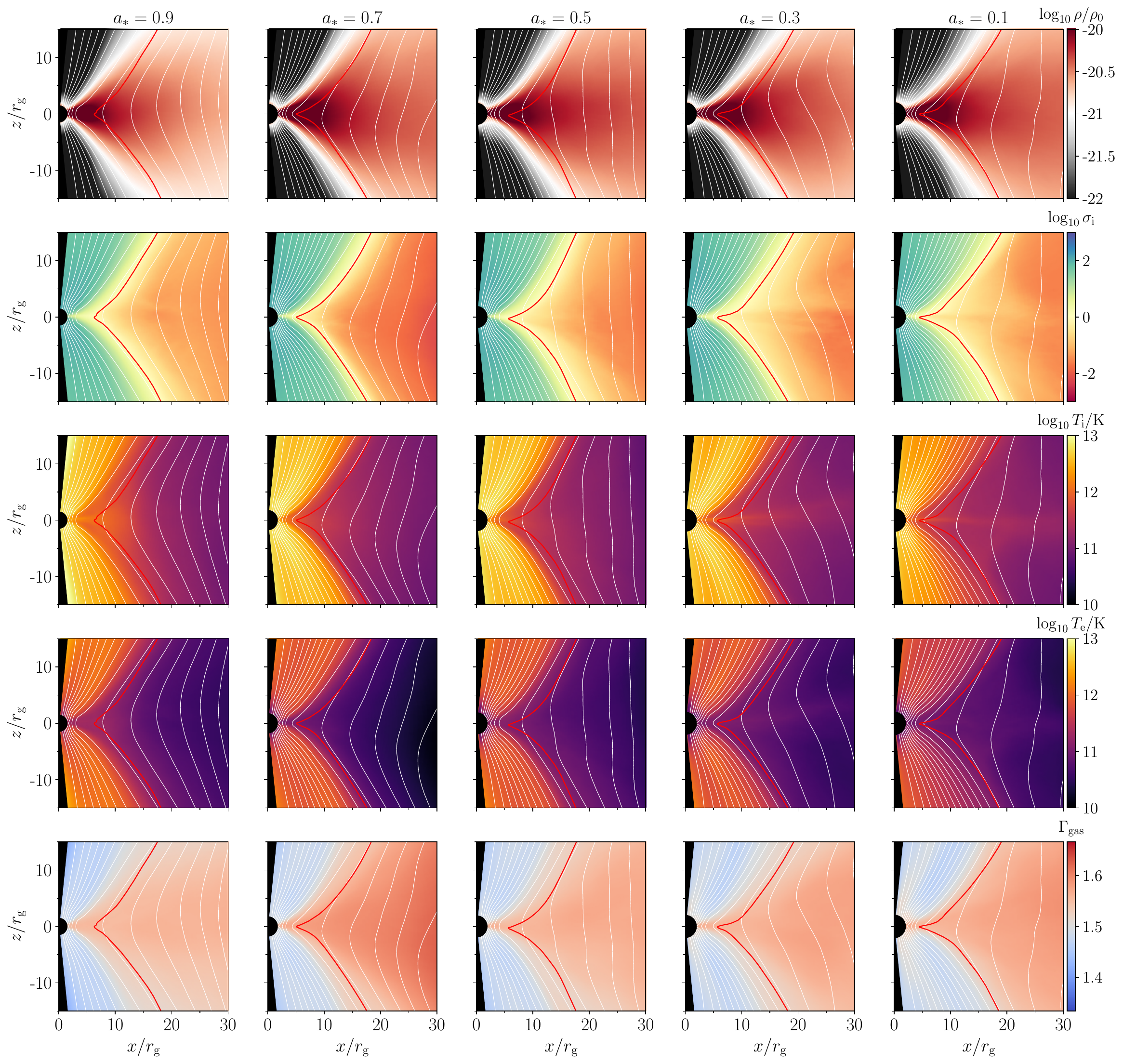}
\caption{Poloidal profiles of prograde radiative simulations. From left to right, we plot poloidal profiles of the time- and azimuth-averaged quantities from the prograde radiative simulations
\texttt{ap9\_radk}, \texttt{ap7\_radk}, \texttt{ap5\_radk}, \texttt{ap3\_radk}, and \texttt{ap1\_radk}. From top to bottom, we plot the averaged rest mass density $\rho/\rho_0$ (where $\rho_0=1$ g cm$^{-3}$), the
magnetization $\sigma_i$,
ion temperature $T_{\rm i}$,
electron temperature $T_{\rm e}$,
and adiabatic index $\Gamma_{\rm gas}$.
All quantities were averaged over $2\pi$ in azimuth and over the final $5000\,t_{\rm g}$ of simulation runtime. Magenta contours indicate $\sigma_{\rm i}=1$; white contours show surfaces of constant azimuthal vector potential $A_\phi$ (\autoref{eq:aphi}).}
\label{fig:phiprofspro}
\end{figure*}

\begin{figure*}
\centering
\includegraphics[width=\textwidth]{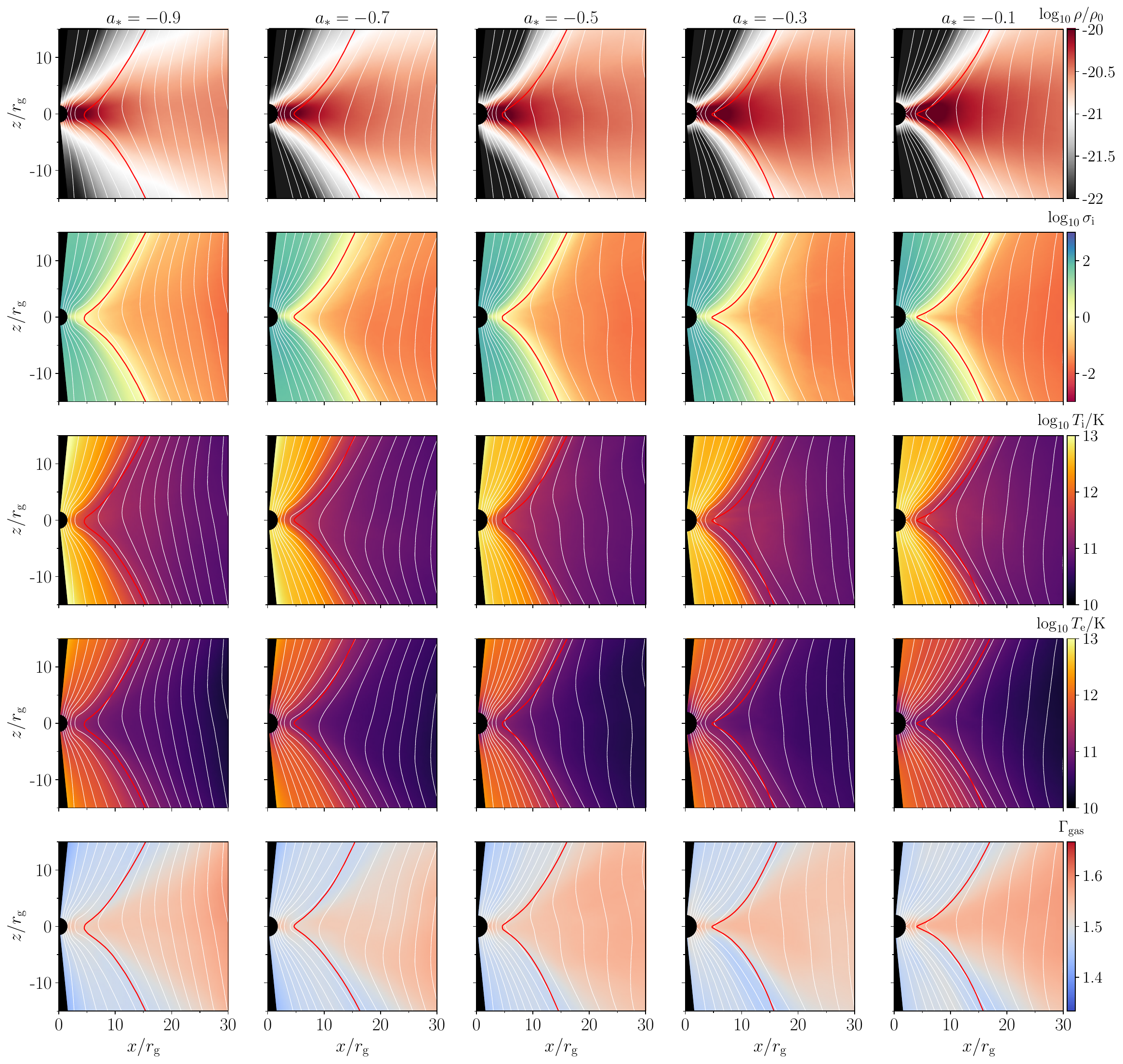}
\caption{Poloidal profiles of retrograde radiative simulations. We plot the same quantities as in \autoref{fig:phiprofspro} for the retrograde radiative simulations \texttt{am9\_radk}, \texttt{am7\_radk}, \texttt{am5\_radk}, \texttt{am3\_radk}, and \texttt{am1\_radk}.
}
\label{fig:phiprofsret}
\end{figure*}

\begin{figure*}
\centering
\includegraphics[width=\textwidth]{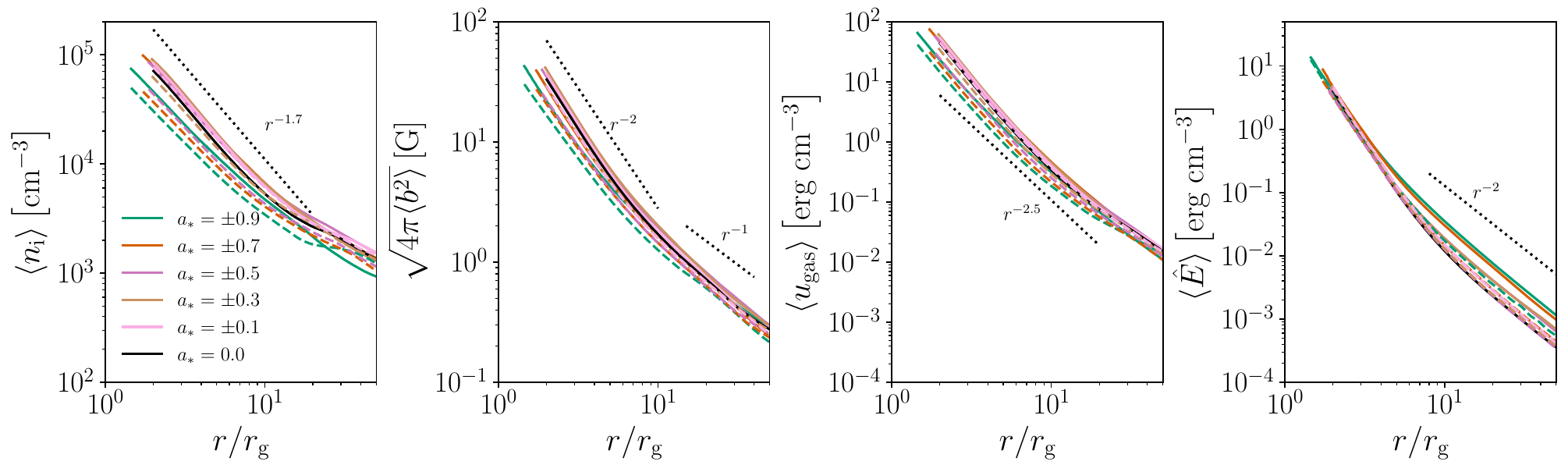}
\caption{Radial profiles of time-averaged simulation quantities. From left to right, we plot radial profiles of the averaged ion number density $\langle n_{\rm i}\rangle$,
the fluid frame magnetic field strength in Gauss, $\sqrt{4\pi\langle b^2\rangle}$, the gas internal energy density $\langle\ugas\rangle$, and the fluid-frame radiation energy density $\langle\hat{E}\rangle$. All quantities are averaged in time, azimuth, and polar angle using \autoref{eq:radavg}. The profiles for prograde simulations are shown in solid lines and profiles for retrograde simulations are shown in dashed lines. Dot-dashed lines indicate characteristic radial falloffs for each quantity.
}
\label{fig:rprofs1}
\end{figure*}

\begin{figure*}
\centering
\includegraphics[width=\textwidth]{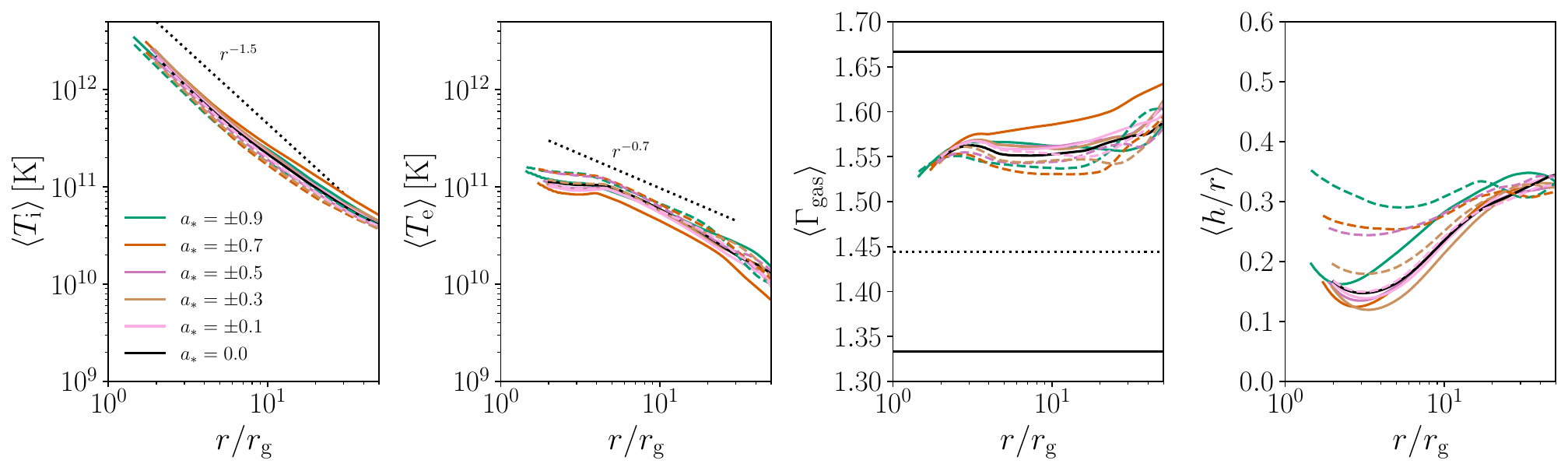}
\caption{More radial profiles of time-averaged simulation quantities. From left to right, we plot radial profiles of the averaged ion temperature $\langle T_{\rm i}\rangle$,
the electron temperature $\langle T_{\rm e}\rangle$, the gas adiabatic index $\langle \adi \rangle$ and the disc scale height $\langle h/r\rangle$. The first three quantities are averaged in time, azimuth, and polar angle using \autoref{eq:radavg}, and the scale height is computed with \autoref{eq:hr}. The profiles for prograde simulations are shown in solid lines and profiles for retrograde simulations are shown in dashed lines. The solid horizontal lines on the third panel indicate the lower and upper limits of $\adi$; $\adi=4/3$ and $\adi=5/3$ for an ultra-relativistic and non-relativistic gas, respectively. The dotted horizontal line indicates $\adi=13/9$.
}
\label{fig:rprofs2}
\end{figure*}

\begin{figure*}
\centering
\includegraphics[width=\textwidth]{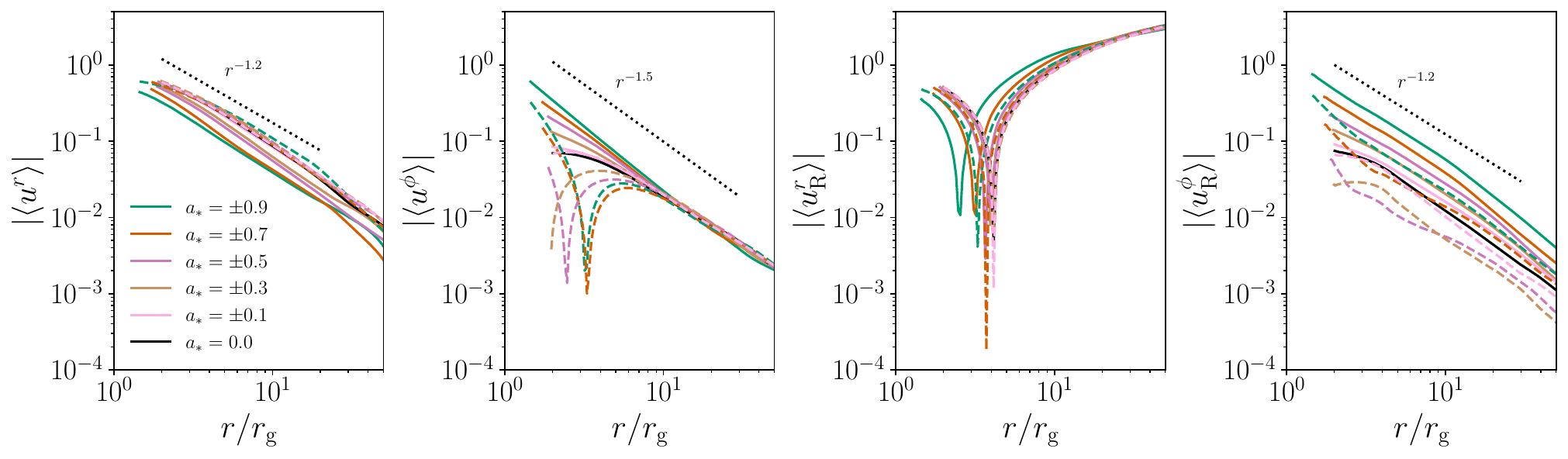}
\caption{Radial profiles of time-averaged simulation velocities. From left to right, we plot radial profiles of the averaged fluid radial velocity $\langle u^r\rangle$,
the averaged fluid azimuthal velocity $\langle u^\phi\rangle$, the averaged radiation frame radial velocity $\langle u^r_{\rm R} \rangle$ and the averaged radiation frame azimuthal velocity $\langle u^\phi_{\rm R}\rangle$. All velocity components are in Kerr-Schild coordinates and are averaged in time, azimuth, and polar angle using \autoref{eq:radavg}. The profiles for prograde simulations are shown in solid lines and profiles for retrograde simulations are shown in dashed lines.
}
\label{fig:rprofs3}
\end{figure*}

\begin{figure*}
\centering
\includegraphics[width=0.7\textwidth]{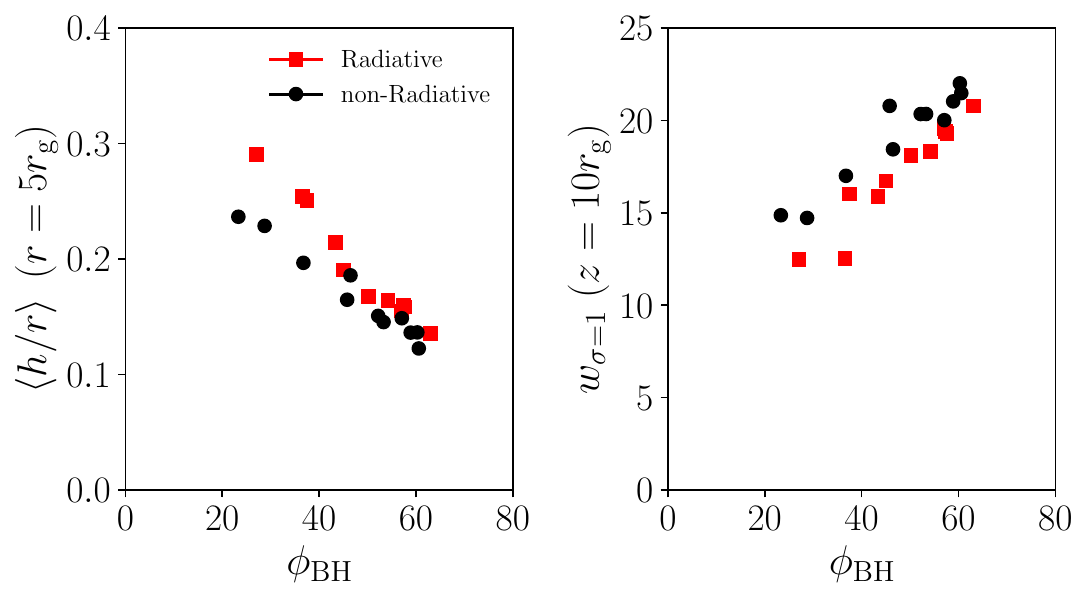}
\caption{Dependence of disc height and jet width on magnetic flux. The left plot shows values of the disc scale height $\langle h/r\rangle$ computed at radius $r=5\,r_{\rm g}$ with \autoref{eq:hr} plotted against the corresponding simulation's averaged value of dimensionless magnetic flux $\phi_{\rm BH}$ (\autoref{eq:madpar}). The right plot shows values of the jet width $w$, defined as twice the cylindrical radius of the average $\sigma_{\rm i}=1$ contour (the magenta lines in \autoref{fig:phiprofspro} and \autoref{fig:phiprofsret}) at height $z=10\,r_{\rm g}$. In both plots, red squares indicate values computed from the eleven radiative simulations and black circles indicate values taken from the eleven non-radiative simulations.
}
\label{fig:hrplot}
\end{figure*}

In \autoref{fig:phiprofspro} and \autoref{fig:phiprofsret}, we display poloidal profiles of certain quantities averaged in time (over $t=15000-20000\,t_{\rm g}$) and azimuth (over $2\pi$) for the prograde and retrograde radiative simulations, respectively.\footnote{We do not show poloidal averages of the $a_*=0$ simulation \texttt{ap0\_radk}, but it is qualitatively similar to the low-spin simulations in \autoref{fig:phiprofspro} and \autoref{fig:phiprofsret}.} In each panel, from top to bottom, we plot the time- and azimuthally-averaged mass density $\rho$, the averaged magnetization parameter $\sigma_{\rm i}$, the ion temperature $T_{\rm i}$, the electron temperature $T_{\rm e}$, and the adiabatic index $\adi$. In all panels we indicate the averaged $\sigma_{\rm i}=1$ surface in a magenta contour. To indicate average poloidal magnetic field lines, we also plot contours of the average phi-component of the vector potential, computed by integrating the radial magnetic field $B^r$ on contours of constant $r$:
\begin{equation}
\label{eq:aphi}
    A_\phi=\int_\theta B^r \sqrt{-g} \ed \theta,
\end{equation}
where in the integral we use the time- and azimuthally-averaged radial field $B^r$ in Kerr-Schild coordinates.

The poloidal profiles in \autoref{fig:phiprofspro} and \autoref{fig:phiprofsret} indicate that all the radiative simulations have average structures close to the black hole that are characteristic for hot magnetically arrested discs. Each simulation features thick discs with a large scale height and wide jets, as traced either by the $\sigma_{\rm i}=1$ surface or the last $A_\phi$ contour to thread the black hole. The jet region is strongly magnetized, with $\sigma_{\rm i}$ reaching the ceiling value of the simulation. While the discs at large radii are weakly magnetized, the average $\sigma_{\rm i}$ exceeds unity out to $r\approx5\,r_{\rm g}$ in the equatorial plane in all simulations. Since the 230 GHz emission region probed by the EHT extends to $r\approx5\,r_{\rm g}$ \citep{PaperV}, accurate emission modelling of high-magnetization material is essential for interpreting EHT observations.

The ion and electron temperature plots in the third and fourth in \autoref{fig:phiprofspro} and \autoref{fig:phiprofsret} indicate that in the inner $30\,r_{\rm g}$ of the simulation both electrons and ions are hot, $10^{10}\mathrm{K}<T_{\rm i,e}<10^{13}\mathrm{K}$, with both electron and ion temperatures climbing to their highest values in the nearly evacuated jet region and decreasing with radial distance in the equatorial plane.
The average electron temperature is less than the average ion temperature throughout the simulation domain, despite the fact that in the $\texttt{K19}$ heating prescription (\autoref{eq::kawazura}) delivers most of the heat to electrons in the most highly magnetized regions ($\delta_{\rm e}>0.5$ when $\beta_{\rm i}\ll1$). This result suggests that radiation plays a key role in these simulations in keeping the electrons cooler than the ions, even in the highly magnetized jet. The last rows of \autoref{fig:phiprofspro} and \autoref{fig:phiprofsret} indicate that the variation of $T_{\rm i,e}$ over the simulation volume results in a combined gas adiabatic index $\adi$ that is not constant. In particular, in the hot jet region electrons become ultra-relativistic and ions are near-relativistic, and the adiabatic index $\adi$ decreases from its value in the disc closer to the relativistic limit of $4/3$.

In \autoref{fig:rprofs1}, \autoref{fig:rprofs2}, and \autoref{fig:rprofs3} we show radial profiles of certain density-weighted quantities for the prograde (in solid lines) and retrograde (in dashed lines) radiative simulations. For a simulation quantity $q$, we define a density-weighted average $\langle q \rangle(r)$ over the $(\theta,\phi,t)$ coordinates as:

\begin{equation}
\label{eq:radavg}
\langle q \rangle(r) = \frac{\int_t\int_\phi\int_\theta q \rho \sqrt{-g}\;\ed\phi \ed\theta \ed t}{\int_t\int_\phi\int_\theta \rho \sqrt{-g}\;\ed\phi \ed\theta \ed t}.
\end{equation}

In \autoref{fig:rprofs1} we plot the averaged ion number density $\langle n_{\rm i}\rangle$, the averaged fluid-frame magnetic field strength in Gauss $\sqrt {4\pi\langle b^2 \rangle}$, the average fluid internal energy density $\langle \ugas \rangle$ and the average radiation energy density in the fluid frame $\langle \hat{E}\rangle$. We find that the radiative simulations at all values of black hole spin show similar power-law falloffs for these disc-averaged quantities. The density falls off as ${\approx}r^{-1.7}$, though the retrograde simulations exhibit lower densities and slightly shallower falloffs. At larger radii the magnetic field strength falls off as ${\approx} r^{-1}$, but at small radii $r<10\,r_{\rm g}$ the magnetic field strength falls off more steeply as ${\approx} r^{-2}$; this change in the slope of $b(r)$ indicates that the disc magnetic field becomes poloidally dominated close to the black hole, which is a key signature of magnetically arrested accretion. The gas internal energy density $\ugas$ falls off as ${\approx} r^{-2.5}$ in the inner disc, but the retrograde discs tend to have slightly lower values. Finally, the radiation energy density $\hat{E}$ has a falloff consistent with free streaming ${\approx} r^{-2}$ outside of $r=10\,r_{\rm g}$ but falls off more steeply interior to this radius, indicating that most of the bolometric luminosity is produced in the inner region of the accretion flow.

In \autoref{fig:rprofs2} we plot the density-weighted average ion and electron temperatures $\langle T_{\rm i} \rangle$, $\langle T_{\rm e} \rangle$ in Kelvin, the average gas adiabatic index $\langle \adi \rangle$, and the disc scale height $\langle h/r \rangle$, defined as
\begin{equation}
\label{eq:hr}
\langle h/r \rangle(r) = \frac{\int_t\int_\theta\int_\phi \left|\pi/2-\theta\right| \rho \sqrt{-g}\;\ed\theta \ed\phi \ed t}{\int_t\int_\theta\int_\phi \rho \sqrt{-g}\;\ed\phi \ed\theta \ed t}.
\end{equation}
Again, we see that the average value of the electron temperature is less than the averaged ion temperature in the inner disc, though the two species have different temperature profiles, with the ion temperature falling off more steeply compared to the relatively flat electron temperature profile for $r<10\,r_{\rm g}.$ The retrograde simulations tend to have slightly cooler ions and slightly hotter electrons than the corresponding prograde simulations, though this spin dependence is mild.

The third panel of \autoref{fig:rprofs2} shows the average value of the combined adiabatic index $\adi$ computed with \autoref{eq::gammaeff}. Throughout the inner disc $r<50\,r_{\rm g}$, the radiative simulations produce values of $\adi$ between the non-relativistic limit $\adi=5/3$ and the equal-temperature value for relativistic electrons and non-relativistic ions $\adi=13/9$; the average $\adi\approx1.55$. The retrograde simulations have slightly hotter (and thus more relativistic) electrons; correspondingly, their average adiabatic indices are slightly lower than their prograde counterparts. The $a_*=0.7$ simulation stands out with a value of $\adi$ elevated by ${\approx} 20\%$ compared to the other prograde simulations; this seems to be the result of the electrons in this simulation being slightly cooler in comparison to the $a_*=0.5$ and $a_*=0.9$ simulations, though the reason for this systematic difference is not immediately clear. The disc scale height is strongly spin dependent, with retrograde simulations showing larger values of $\langle h/r \rangle$ in the inner disc where the fluid transitions from counter-rotating to co-rotating with the black hole. The retrograde simulations with more rapidly spinning black holes have the largest values of $\langle h/r\rangle$; this result is consistent with what was observed in \citet{Narayan22}.

In \autoref{fig:rprofs3} we plot the absolute value of the density-weighted averaged radial and azimuthal 4-velocities $\langle u^r \rangle$ and $\langle u^\phi\rangle$ in Kerr-Schild coordinates. We also plot the corresponding 4-velocity components for the radiation frame $\langle u^r_{\rm R}\rangle$, $\langle u^\phi_{\rm R}\rangle$. The fluid's radial velocity increases toward the black hole with an ${\approx} r^{-1.2}$ dependence; retrograde simulations have a larger inflow velocity than prograde simulations. The angular velocity has a Keplerian fall-off ${\approx} r^{-1.5}$, though with slightly sub-Keplerian values of $u^\phi$. The retrograde simulations all show a reversal in the sign of $u^\phi$ at a spin-dependent radius; the more rapidly spinning black holes in retrograde simulations force the fluid to co-rotate at larger radii.

The radial component of the radiation-frame four-velocity changes sign in all simulations in the inner $5\,r_{\rm g}$, indicating that the bulk of the radiation in the inner accretion disc falls into the black hole while farther out radiation escapes to infinity (this does not indicate that no radiation from the inner $5\,r_{\rm g}$ escapes, as \autoref{fig:rprofs3} shows only the density-weighted average of $u_{\rm R}^r$). The azimuthal component of the radiation frame velocity is highly spin-dependent, with more rapidly spinning prograde and retrograde black holes producing larger values of $|u_{\rm R}^\phi|$ and thus radiation that is more highly beamed than black holes with lower spin.

Comparing \autoref{fig:phiprofspro} and \autoref{fig:phiprofsret}, we can see by eye that the jets in the prograde simulations (marked by the $\sigma_{\rm i}=1$ contours) are somewhat wider than in the corresponding retrograde simulations; this is a result of the dependence of the jet width and disc scale height $\langle h/r\rangle$ on the magnetic flux $\phi_{\rm BH}$ noted in \citet{Narayan22}.
In \autoref{fig:hrplot} we compare the value of the disc scale height $\langle h/r\rangle$ at $r=4\,r_{\rm g}$ as a function of the dimensionless horizon flux $\phi_{\rm BH}$ for the radiative and non-radiative simulations. We also show the jet width for all simulations as a function of $\phi_{\rm BH}$, where we define the jet width $w$ as twice the cylindrical radius of the time- and phi-averaged $\sigma=1$ at height $z=10\,r_{\rm g}$.

In \autoref{fig:hrplot} we reproduce the qualitative trend of \citet{Narayan22} that the disc scale height decreases with the magnetic flux $\phi_{\rm BH}$ and the jet width correspondingly increases. The radiative simulations systematically have larger scale heights and narrower jets than their non-radiative counterparts, with the radiative discs (jets) being ${\approx}$15\% thicker (${\approx}10\%$ narrower) than their non-radiative counterparts. This systematic offset is likely a result of the difference in the adiabatic index $\adi$ between the radiative simulations, where $\adi$ is solved for self-consistently and takes on a disc-averaged value $\adi{\approx}1.55$, compared with the non-radiative simulations, where $\adi$ is fixed at $\adi=13/9$. The larger effective adiabatic index in the radiative simulation implies that radiative discs with similar internal energy densities to their non-radiative counterparts will have ${\approx} 25\%$ more gas pressure; this systematically increased pressure in the radiative simulations likely increases the scale height and decreases the width of the magnetized funnel. Naively, we may have expected that the discs in the radiative simulations would be thinner than the discs in the corresponding non-radiative simulations due to an overall loss of internal energy density from cooling; however, because the radiative efficiency is low $(3-35\%$), the larger difference in the adiabatic indices from self-consistently tracking the electron and ion species in the radiative simulations has a larger effect than radiative cooling in modifying the gas pressure.

\subsection{Temperature Ratio}
\label{sec:rbeta}

\begin{figure*}
\centering
\includegraphics[width=0.7\textwidth]{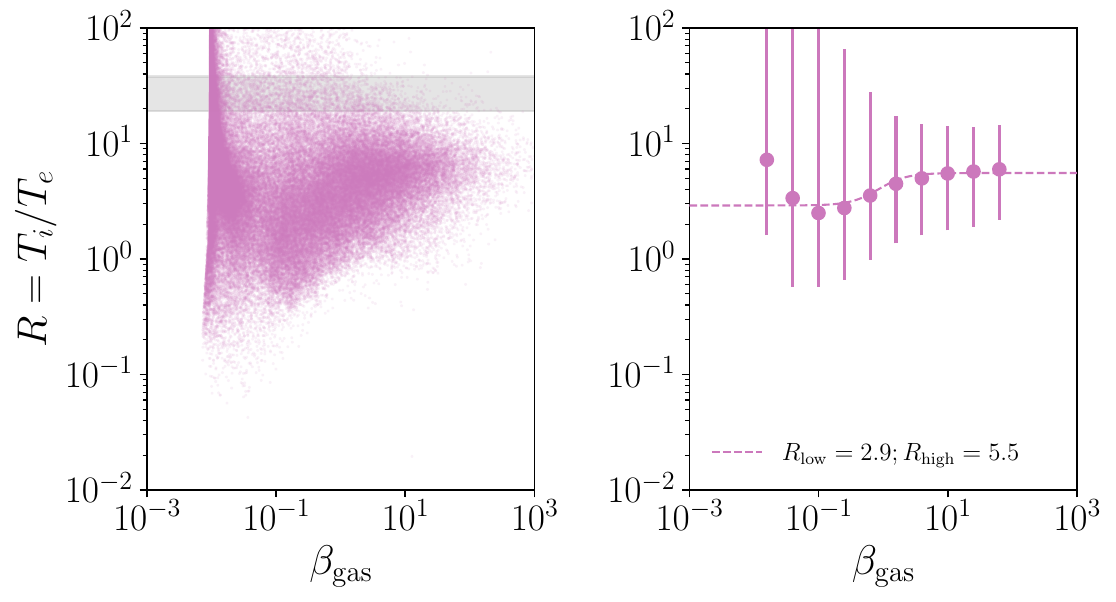}
\caption{Temperature ratio $R=T_{\rm i}/T_{\rm e}$ versus $\beta_{\rm gas}$ in simulation \texttt{ap5\_radk}. In the left panel, we show a scatter plot of $R$ against $\beta_{\rm gas}$ with points sampled randomly from all cells within $r<25\,r_{\rm g}$ over the time range $15000-20000\,t_{\rm g}$ in the simulation.
We exclude cells with $\sigma_{\rm i}>25$, as these cells are also excluded in radiative transfer for the 230 GHz images reported in \autoref{sec:images}.
The initial range of $R$ when the simulation is re-started from single-fluid GRMHD data at $t=10000\,t_{\rm g}$ is indicated with the gray band. In the right panel, we plot the left panel data binned in 10 equally spaced logarithmic bins; data points indicate the median value in each bin and error bars indicate the 68\% credible interval. We fit the binned data to the \citet{Moscibrodzka16} prescription \autoref{eq:rbeta}; the resulting fit for parameters $R_{\rm low}=2.5$, $R_{\rm high}=5.6$ is indicated by the dashed curve.
}
\label{fig:rbetasli}
\end{figure*}

Next we investigate the electron-to-ion temperature ratio $R=T_{\rm i}/T_{\rm e}$  in individual snapshots of our radiative, two-temperature simulations. To simulate horizon-scale synchrotron images from single-fluid GRMHD simulations without self-consistent temperature evolution, the standard approach \citep[e.g.][]{PaperV,SgrAPaperV,PaperVIII} is to use the \citet{Moscibrodzka16} phenomenological function $R(\beta_{\rm gas})$:
\begin{equation}
\label{eq:rbeta}
R = R_{\rm low}\frac{1}{1+\beta^2_{\rm gas}} + R_{\rm high}\frac{\beta^2_{\rm gas}}{1+\beta^2_{\rm gas}}.
\end{equation}
In the \citet{Moscibrodzka16} ``$R-\beta$'' model (\autoref{eq:rbeta}), the ion-to-electron temperature ratio is a fixed function of the gas plasma beta parameter and transitions from a value $R=R_{\rm low}$ in highly magnetized regions $\beta_{\rm gas}\ll 1$ to $R=R_{\rm high}$ in weakly magnetized regions $\beta_{\rm gas}\gg 1$. Because electron heating functions from plasma turbulence simulations \citep{Howes10,Kawazura19} preferentially heat electrons at low $\beta_{\rm gas}$ and preferentially heat ions at high $\beta_{\rm gas}$, libraries of simulated 230 GHz images used for interpreting EHT results tend to adopt $R_{\rm low}=1$ and sample several values $R_{\rm high}>1$ \citet{PaperV}, though in simulations of M87* \citet{PaperVIII,PaperIX} also explored models with $R_{\rm low}=10$ to simulate the effect of radiative cooling of electrons in high magnetization regions.

With our radiative, two-temperature simulation suite with electrons and ions heated by the \citet{Kawazura19} plasma turbulence prescription, we can investigate the dependence of $R$ against $\beta_{\rm gas}$ and compare it to the standard \citet{Moscibrodzka16} prescription for single-fluid GRMHD, \autoref{eq:rbeta} \citep[see also][]{Dihingia23}.
In the left panel of \autoref{fig:rbetasli}, we show a scatter plot of the electron-to-ion temperature ratio versus the plasma beta parameter $\beta_{\rm gas}$ from randomly chosen cells in the inner $25 \,r_{\rm g}$ of the simulation \texttt{ap5\_radk} in the time range $15000-20000\,t_{\rm g}$. The gray shaded region in the left panel of \autoref{fig:rbetasli} indicates the range of initial temperature ratios $R_{\rm init}=19-38$ set when we initialize the two-temperature plasma.\footnote{With $u_{\rm e}=0.05 \ugas$, $u_{\rm i}=0.95\ugas$; $R_{\rm init}=19$ in cold regions where electrons are non-relativistic ($\Gamma_{\rm e}=5/3$), and $R_{\rm init}=38$ in hot regions where electrons are relativistic ($\Gamma_{\rm e}=4/3$), assuming $\Gamma_{\rm i}=5/3$.}
The range of $\beta_{\rm gas}$ in the simulation is limited by an overall floor $b^2/u_{\rm gas}<200$ used in \texttt{KORAL} along with the ceiling $\sigma_{\rm max}=100$ to ensure numerical stability in highly magnetized regions.

In the left panel of \autoref{fig:rbetasli} we can see that in our two-temperature simulations, the temperature ratio $R$ typically takes on intermediate values $R\approx 1-10$, except for in the most magnetized regions close to the $\beta_{\rm gas}$ floor where radiative cooling is strongest. Furthermore, we see that $R$ is not a single-valued function of $\beta_{\rm gas}$;
there
can be up to an order of magnitude of scatter in $R$ for simulation cells with a given $\beta_{\rm gas}$. The scatter in the $R-\beta_{\rm gas}$ relationship arises because the electron and ion temperatures of a packet of gas depend not only on the local heating fraction $\delta_{\rm e}$ through \autoref{eq::kawazura}, but also on the thermal history of the gas packet through transport, adiabatic compression/expansion, and radiation.

In the right panel of \autoref{fig:rbetasli} we bin the $R$ values in 10 equally spaced logarithmic bins ranging from $\beta_{\rm gas}=10^{-2}$ to in $\beta_{\rm gas}=10^2$ and plot the resulting median values and $68\%$ credible interval for each bin.
We fit the two-parameter \citet{Moscibrodzka16} model, \autoref{eq:rbeta}, to the binned data, resulting in a model with $R_{\rm low}=2.5$, $R_{\rm high}=5.6$. While we only show results for model \texttt{ap5\_radk} here, the distributions of $R$ and the fitted values of $R_{\rm low}$, $R_{\rm high}$ for the simulations with different black hole spin values are similar; we summarize the fitted $R_{\rm low}$, $R_{\rm high}$ values for all the radiative simulations in \autoref{tab:rbeta}.

\begin{table}
    \centering
    \begin{tabular}{l|ll}
        \toprule
        Model & $R_{\rm low}$ & $R_{\rm high}$ \\
        \midrule
        \texttt{ap9\_radk} & 3.2&5.4 \\
        \texttt{ap7\_radk} & 3.5&8.1 \\
        \texttt{ap5\_radk} & 2.5&5.6 \\
        \texttt{ap3\_radk} & 2.8&5.6 \\
        \texttt{ap1\_radk} & 2.2&5.5 \\
        \texttt{a0\_radk}  & 1.8&5.7 \\
        \texttt{am1\_radk} & 2.1&4.8 \\
        \texttt{am3\_radk} & 1.6&5.0 \\
        \texttt{am5\_radk} & 2.2&4.6 \\
        \texttt{am7\_radk} & 1.6&4.2 \\
        \texttt{am9\_radk} & 2.2&4.6 \\
        \bottomrule
    \end{tabular}
    \caption{\citet{Moscibrodzka16} electron temperature parameters $R_{\rm low}$,$R_{\rm high}$ fit to binned radiative simulation data for all eleven radiative simulations. For each simulation we bin raw $R,\beta_{\rm i}$ values from simulation snapshots and fit \autoref{eq:rbeta} as described in \autoref{fig:rbetasli}.
    }
    \label{tab:rbeta}
\end{table}
Our radiative, two-temperature simulations heated by sub-grid plasma turbulence prefer moderate ion-to-electron temperature ratios, summarized by moderate values of the fitted \citet{Moscibrodzka16} parameters $R_{\rm low}\approx2$, $R_{\rm high}\approx5$. In our simulations, close to the black hole electrons are only moderately cooler than ions, even with strong radiative cooling. This moderate temperature ratio is consistent with results for 2D radiative two-temperature simulations at similar accretion rates heated by the \citet{Howes10} electron heating prescription in \citet{Dihingia23}. Notably, most single fluid MAD GRMHD models that satisfy EHT polarimetric constraints of M87* have higher values $R_{\rm high}\gtrsim80$, resulting in an overall much cooler population of electrons \citep{PaperVIII}.

\subsection{Images}
\label{sec:images}

\begin{figure*}
\centering
\includegraphics[width=\textwidth]{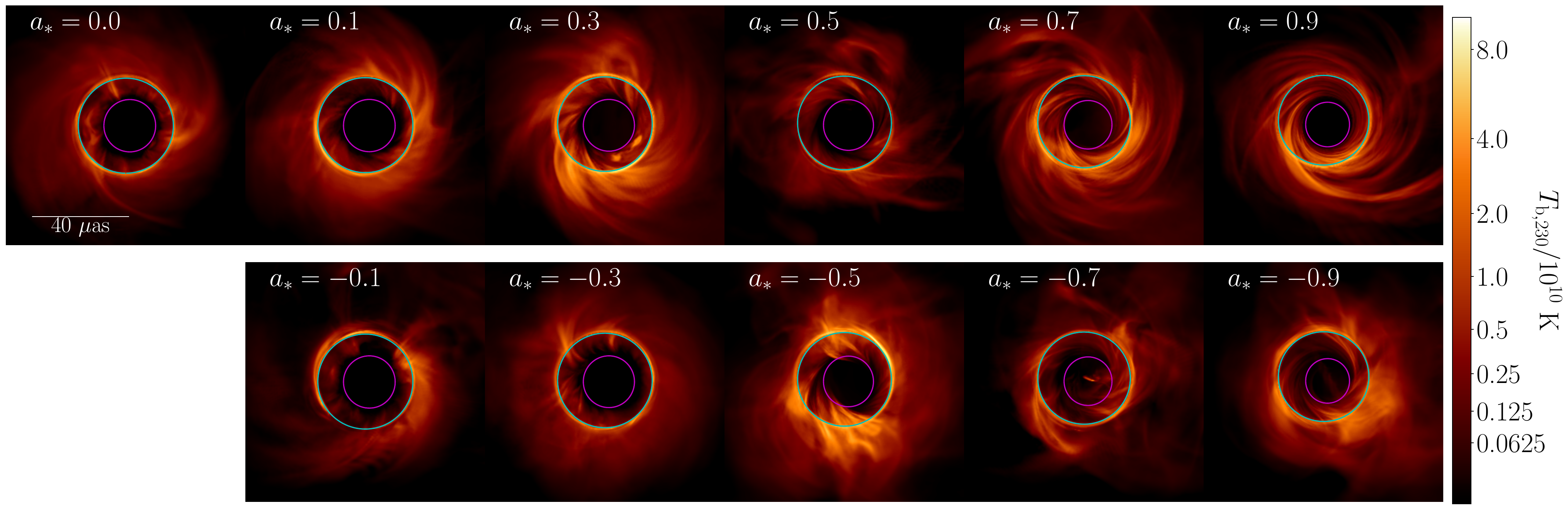}
\caption{Snapshot images from the eleven radiative simulations. Synchrotron images were produced at 230 GHz using \texttt{ipole}; prograde simulations (top row) have an observer inclination $\theta_{\rm o}=163\deg$ and retrograde simulations (bottom row) have an observer inclination $\theta_{\rm o}=17\deg$, such that in all simulations the black hole spin is oriented $163\deg$ from the line of sight; the spin vector points to the East (left) of the image \citep{Mertens2016}. Images are plotted in units of brightness temperature $T_{\rm b}$ in a gamma colour scale with image intensities $I$ scaled as $I^{1/4}$. The cyan curve in each image is the ``critical curve'' that indicates the position of the multiply-lensed photon ring \citep{Johnson20}; the magenta curve denotes the direct image of the event horizon, or ``inner shadow'' edge \citep{Chael21}.
}
\label{fig:snapshots}
\end{figure*}

\begin{figure*}
\centering
\includegraphics[width=\textwidth]{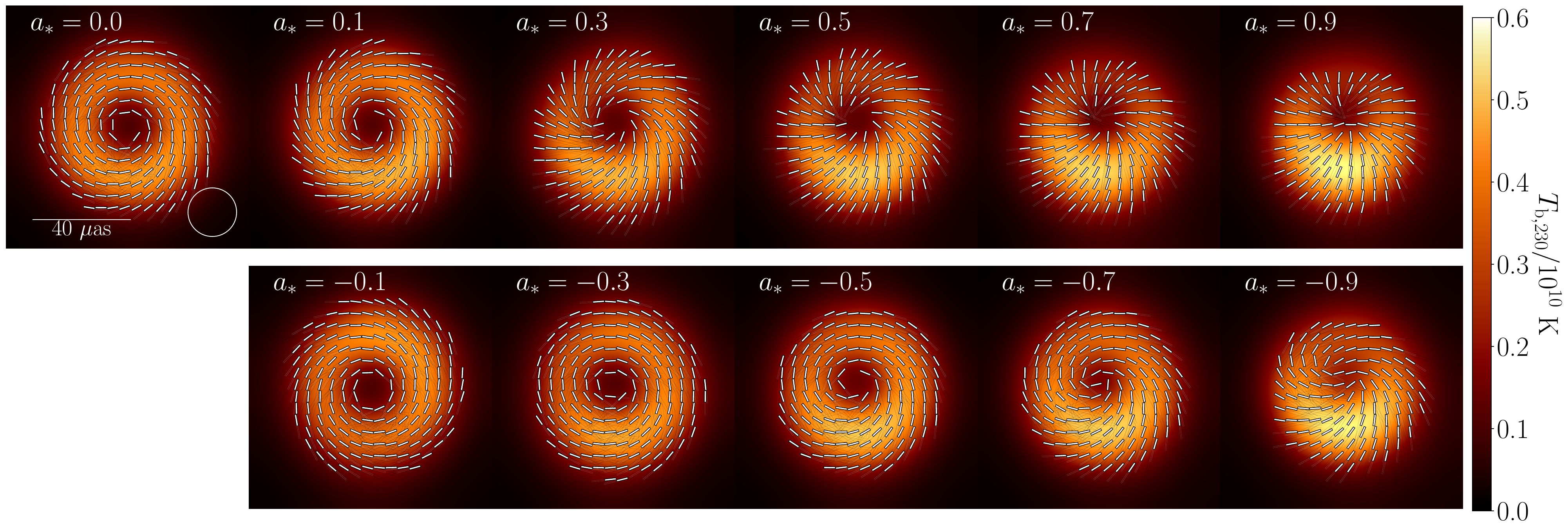}
\caption{Averaged images from the eleven radiative simulations blurred to EHT resolution. Images from each simulation were blurred to the EHT $20\,\mu$as resolution with a circular Gaussian kernel (indicated with the white circle in the upper left image) and time averaged over the range $15000-20000\,t_{\rm g}.$ Images are displayed in a linear scale, and ticks indicate the EVPA direction across the image.
}
\label{fig:avgs}
\end{figure*}

\begin{figure*}
\centering
\includegraphics[width=\textwidth]{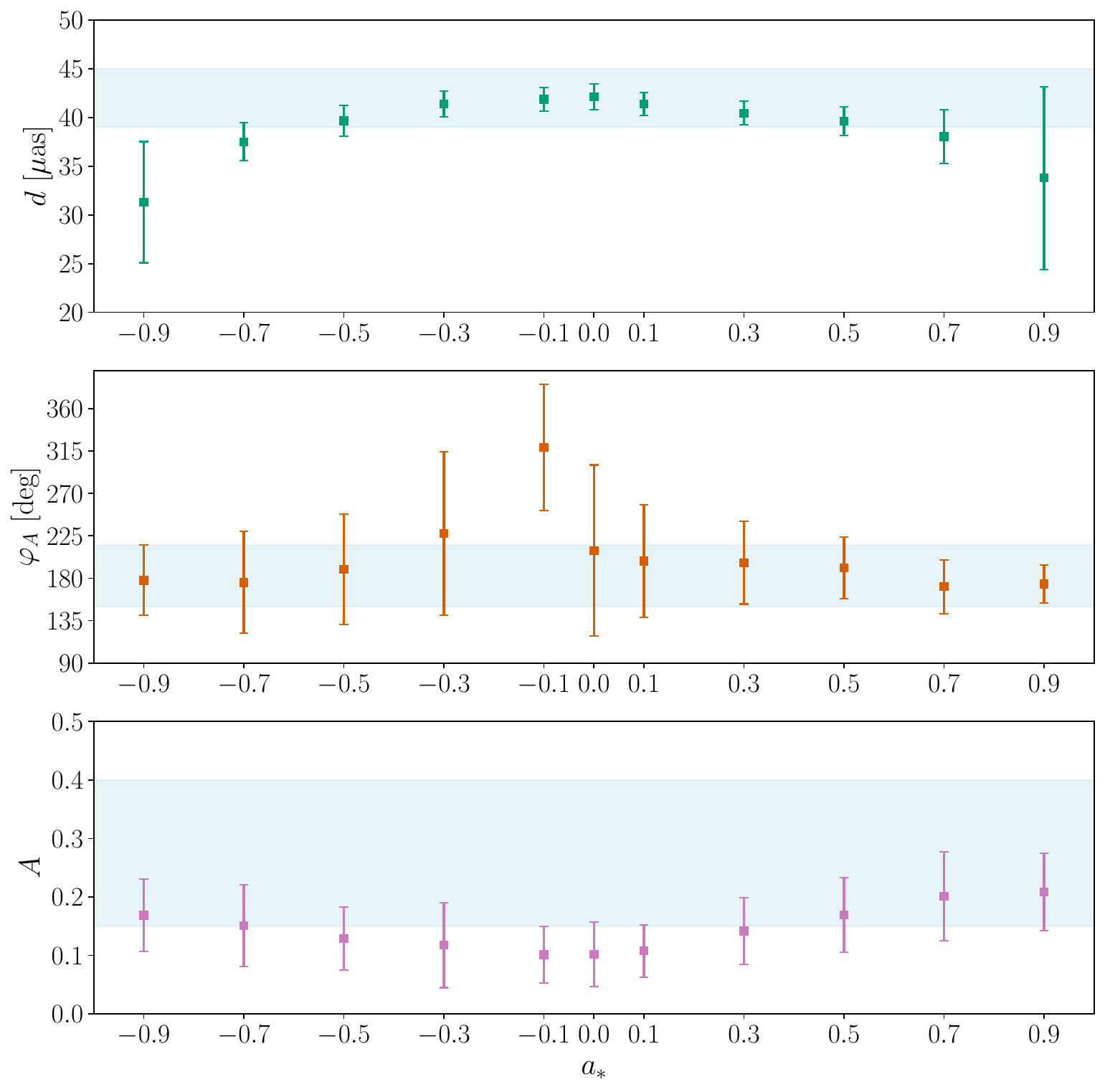}
\caption{Total intensity image statistics. From top to bottom we plot the ring diameter $d$ (\autoref{eq:diam}), image position angle $\varphi_A$ (\autoref{eq:posang}) and image asymmetry parameter $A$ (\autoref{eq:asym}) for each radiative simulation as a function of spin $a_*$. We plot the mean value and $1\sigma$ error bars from snapshots in the range $15000-20000\,t_{\rm g}$. The blue shaded ranges indicate the range of values measured from EHT observations of M87* in 2017 \citet{PaperIV} and 2018 \citet{EHT2024M87}.
}
\label{fig:Istats}
\end{figure*}

\begin{figure*}
\centering
\includegraphics[width=\textwidth]{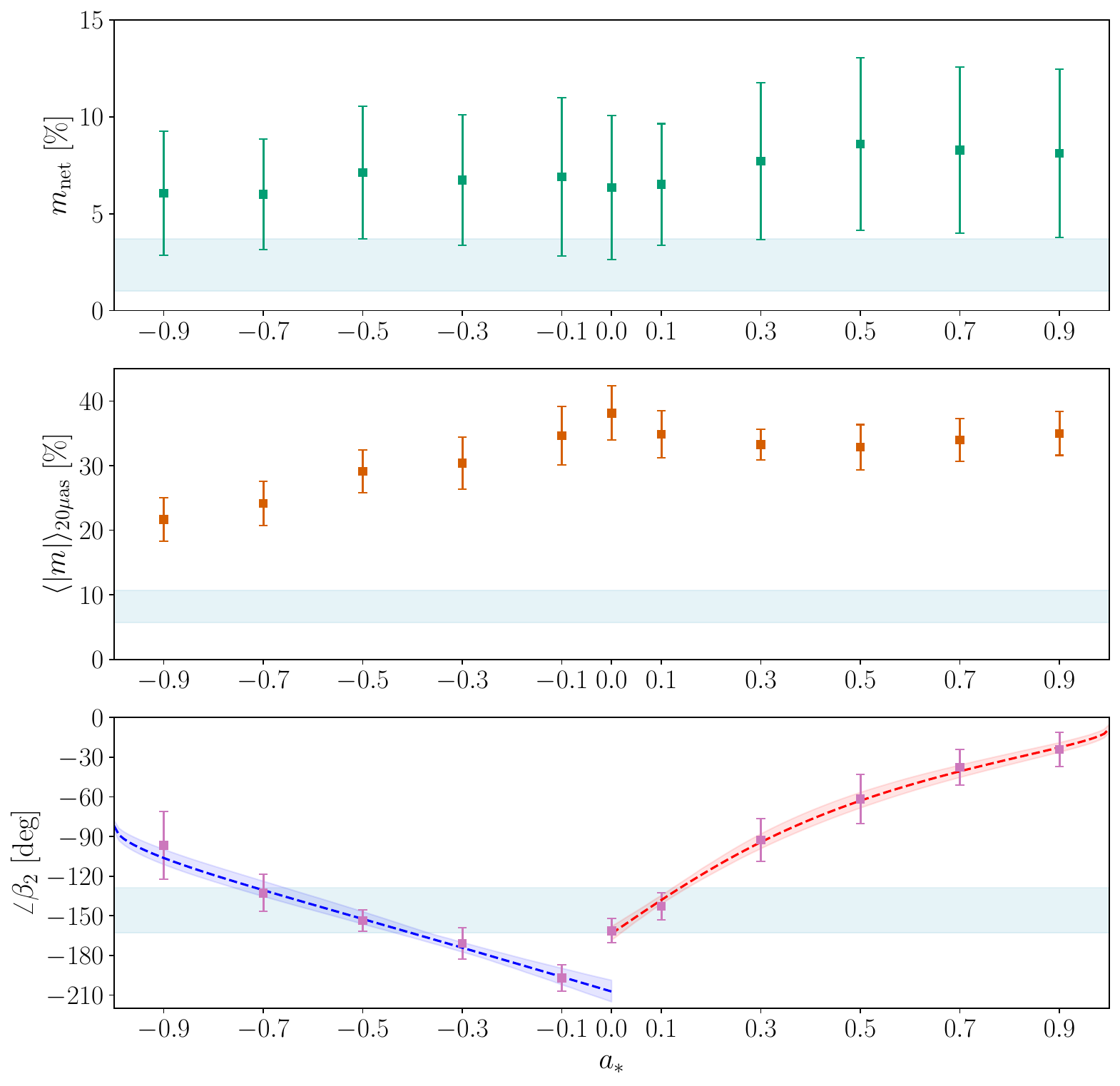}
\caption{Polarimetric image statistics. From top to bottom we plot the net linear polarization fraction $m_{\rm net}$ (\autoref{eq:mnet}), the EHT-scale average linear polarization fraction $\langle|m|\rangle$ (\autoref{eq:mavg}) and the phase of the linear polarization's second Fourier coefficient $\angle \beta_2$ (\autoref{eq:Beta2}, \citealt{Palumbo20}). We plot the mean value and $1\sigma$ error bars from snapshots in the range $15000-20000\,t_{\rm g}$. The blue shaded ranges indicate the range of values measured from EHT observations of M87* in 2017 \citet{PaperVII}. The dashed blue and red lines indicate fits to the prograde and retrograde simulation results for $\angle\beta_2$ as a function of $a_*$ using the functional form \autoref{eq:fit}. The shaded blue and red regions represent the $99\%$ credible interval on the fitting functions for the prograde and retrograde simulations derived from sampling the fit covariance.
}
\label{fig:polstats}
\end{figure*}

In \autoref{fig:snapshots} we show $230$ GHz image snapshots selected from all eleven radiative simulations computed with \texttt{ipole} \citep{ipole}. The observer inclination in our simulated images is set to $\theta_{\rm o}=163\deg$ for the prograde discs and $\theta_{\rm o}=17\deg$ for the retrograde discs; the images are plotted in a gamma colour scale with intensity $I$ scaled as $I^{1/4}$. The snapshot images in \autoref{fig:snapshots} show similar intensity distributions to images from single-fluid GRMHD MAD simulations known to fit total intensity observations of M87* \citep{PaperV}. Notably, the optically thin photon ring feature from light rays lensed $>180\deg$ \citep{Johnson20} is apparent in each simulation (indicated by the cyan contour); the ``inner shadow'' feature, or the direct image of material just outside the event horizon (\citealt{Chael21}, indicated by the magenta contour), is also visible.

In \autoref{fig:avgs} we show the $230$ GHz images time-averaged over the interval $15000-20000\,t_{\rm g}$ and blurred to the EHT resolution of 20 $\mu$as with a circular Gaussian kernel. The time-averaged images are displayed in a linear colour scale; we also display the averaged pattern of linear polarization with tick marks indicating the direction of the electron vector position angle (EVPA) across the image. It is apparent in the averaged EHT resolution images that the direction and magnitude of the asymmetry in the image is spin-dependent, with images from more rapidly spinning black holes exhibiting more pronounced brightness asymmetry
\citep{PaperV, SgrAPaperV, Medeiros22}. It is also apparent that the pattern of linear polarization is highly spin-dependent, with lower-spin images displaying more azimuthal polarization patterns and more rapidly spinning black holes displaying more radial polarization patterns \citep{Palumbo20,PaperVIII}.

For each blurred image snapshot, we follow \citet{PaperIV} and compute several summary statistics for both total intensity and linear polarization. We work in polar coordinates $(\rho,\varphi)$ in the image plane.\footnote{We fix $\rho=0$ at the black hole position; real images must be centred before computing summary statistics, but this centring procedure has only a small effect on simulated images \citep{PaperV,PaperVIII}.
} We compute the ring diameter $d$
by finding the radius $\rho_{\rm pk}$ of peak intensity in radial profiles $I(\rho)$ at fixed image position angles and averaging over $\varphi$:
\begin{align}
\label{eq:diam}
d&=2\left\langle \rho_{\rm pk}(\varphi)
\right\rangle_{\varphi\in\left[0,2\pi\right]}.
\end{align}
Similarly, the ring width $w$ is the mean full width at half maximum (FWHM) value of the individual profiles.
With $d$ and $w$ defined, the inner ring radius is $\rho_{\rm in}=(d-w)/2$ and the outer ring radius is $\rho_{\rm out}=(d+w)/2$. We define the ring asymmetry factor $A$ and position angle $\varphi_A$ as
\begin{align}
    \label{eq:asym}
    A &= \left\langle \frac{\left| \int_\varphi I(\rho, \varphi) e^{i\varphi} \ed \varphi \right| }{\int_\varphi I(\rho,\varphi) \ed \varphi}
    \right\rangle_{\rho \in \left[\rho_{\rm in}, \rho_{\rm out}\right]}, \\
    \label{eq:posang}
    \varphi_A &= \left\langle \arg\left[ \int_\varphi I(\rho, \varphi) e^{i\varphi} \ed \varphi \right]
    \right\rangle_{\rho \in \left[\rho_{\rm in}, \rho_{\rm out}\right]}.
\end{align}
The image position angle $\varphi$ is measured in degrees East of North (counter-clockwise from the top of the image in \autoref{fig:avgs}).

In \autoref{fig:Istats} we plot the mean values of $d,A$, and $\varphi_A$ for each simulation, along with their $1\sigma$ error bars computed from snapshots over the time range $15000-20000\,t_{\rm g}$. For the position angle $\varphi_A$ we use the circular mean and standard deviation. We compare our simulation results for these quantities to conservative ranges derived from EHT observations of M87* reported in
\citet{PaperIV, EHT2024M87}, which we indicate in the blue bands in \autoref{fig:Istats}. In particular, we take the EHT ranges as $d\in \left[39,45\right]\mu$as, $A\in\left[0.15,0.4\right]$, and $\varphi_A\in\left[150,215\right]\deg$.\footnote{Note that the observed diameter $d$ and asymmetry $A$ in 20 $\mu$as resolution EHT images was approximately constant between observations in 2017 reported in \citet{PaperIV} and observations in 2018 reported in \citet{EHT2024M87}, while the image position angle $\varphi_A$ shifted by roughly $30\deg$. Our range for $\varphi_A$ covers both years.}

Nearly all the radiative, two-temperature simulations naturally satisfy the observed ranges of total intensity summary statistics derived from EHT observations in \citet{PaperIV,EHT2024M87}. At our fixed value of black hole mass $M=6.5\times10^9M_\odot$ and distance $D=16.8$ Mpc, the ring diameter in the simulated images decreases with black hole spin as the projected size of the black hole event horizon shrinks. The $a_*=\pm0.9$ simulations have median image ring diameters smaller than the observed range for M87*, but we do not account for uncertainty in the measurement of the black hole mass or distance here.
The image position angle $\varphi_A$ shows a mild dependence on spin, with more rapidly spinning simulations having slightly smaller values of $\varphi_A$. As seen in \autoref{fig:avgs}, the magnitude of the image asymmetry $A$ is larger for more rapidly spinning black holes, though all our simulations satisfy the observed ranges within $1\sigma$.

Notably, the $a_*=-0.1$ simulation \texttt{am1\_radk} has a position angle offset by $90-180\deg$ from the rest of the retrograde simulations. This indicates that the $a_*=-0.1$ simulation has its asymmetry direction set by by Doppler beaming from the sense of rotation of the retrograde accretion disc, not the spin direction of the black hole; flipping the viewing orientation of the $a_*=-0.1$ simulation from $\theta_{\rm o}=17\deg$ to $\theta_{\rm o}=163\deg$ (so the spin vector points West) shifts the position angle in line with observations. This result indicates that the \citet{PaperV} inference that the asymmetry direction of the near-horizon synchrotron ring image is set by the rotation sense of the black hole and not that of the accreting material only holds for moderately to rapidly spinning black holes with $|a_*|\gtrsim0.3$. The frame-dragging effect in the weakly spinning $a_*=-0.1$ simulation is not powerful enough to reverse the rotation direction of the retrograde flow in the 230 GHz emitting region, so the asymmetry direction in the image in this case is set by the large-scale flow, as it is in the $a_*=0$ simulation.

We next turn to image-averaged linear polarimetric quantities defined in  \citep{PaperVII, PaperVIII}. Working with images blurred to $20\,\mu$as, the net polarization fraction $m_{\rm net}$ and the average linear polarization at the EHT resolution are:
\begin{align}
    \label{eq:mnet}
    m_{\rm net}&=\frac{\left| \int_y \int_x \mathcal{P}(x,y)\ed x \ed y \right|}{\int_y \int_x I(x,y) \ed x \ed y}, \\
    \label{eq:mavg}
    \langle|m|\rangle &=\frac{\int_y \int_x \left|\mathcal{P}(x,y)\right|\ed x \ed y}{\int_y \int_x I(x,y) \ed x \ed y},
\end{align}
Where $x,y$ are image coordinates, $I$ is the image total intensity, and $\mathcal{P}=Q+iU$ is the complex linear polarization derived from the $Q$ and $U$ Stokes parameters.
The $\beta_2$ parameter \citep{Palumbo20} is defined as the complex amplitude of the second Fourier mode around the (blurred) ring:
\begin{align}
    \label{eq:Beta2}
    \beta_2=\frac{\int_\varphi \int_\rho \mathcal{P}(\rho,\varphi)e^{-2i\varphi}\rho\ed\rho\ed\varphi}{\int_\varphi \int_\rho I(\rho,\varphi)\rho\ed\rho\ed\varphi}.
\end{align}

For each radiative simulation, we plot the median values of $m_{\rm net},\langle|m|\rangle$, and the phase $\angle\beta_2$, in \autoref{fig:polstats} along with $1\sigma$ error bars from all snapshots in the time interval $15000-20000\,t_{\rm g}$.
For $\angle\beta_2$, we use the circular mean and standard deviation.
We compare the results from each simulation with the observed ranges from EHT polarimetric observations of M87* reported in \citep{PaperVII}; $m_{\rm net} \in \left[1.0,3.7\right]\%$, $\langle|m|\rangle\in \left[5.7,10.7\right]\%$, $\angle\beta_2\in \left[-163,-129\right]\deg$, indicated by the shaded blue rectangles in \autoref{fig:polstats}.

While our radiative simulations all naturally reproduce the total intensity summary statistics from M87* EHT images, they do not satisfy the observed polarimetric constraints. Most notably, the radiative simulations are too polarized, with linear polarization fractions $\langle |m|\rangle$ measured at EHT scales that are a factor of ${\approx}2-10$ times too large compared with observations. This over-polarization is likely a result of the electrons in our simulations being too hot, as indicated by the moderate values of $R=T_{\rm i}/T_{\rm e}$ measured in \autoref{sec:rbeta}. In \citet{PaperVIII}, 230 GHz single-fluid GRMHD images are mostly depolarized by differential internal Faraday rotation along different photon trajectories across the image; strong Faraday rotation in a turbulent plasma rotates the EVPAs in adjacent pixels by different amounts, causing an overall depolarization when the image is blurred to 20 $\mu$as resolution \citep{Ricarte20}. Faraday rotation is more effective in cold plasmas, with rotation suppressed as $1/T^2_{\rm e}$ \citep{Jones79}. The median values of $\langle |m|\rangle$ for the two-temperature MAD models reported here are roughly consistent with the $R_{\rm low}=1$, $R_{\rm high}=10$ MAD M87* model set reported in \citet{PaperVIII}, Figure 26.
While the EHT has not yet claimed a measurement of $R$ in the emitting region from M87* observations,
passing MAD models for M87* in the EHT simulation library prefer large values of the ion-to-electron temperature ratio $R_{\rm high}\approx80-160$ in order to produce sufficient image depolarization \citep{PaperVIII}. If it is indeed a consequence of Faraday depolarization, the observed large degree of image polarization in M87* \citep{PaperVII} likely requires much cooler electrons than are produced in two-temperature simulations with electrons heated with the \texttt{K19} turbulent heating prescription.

The quantity $\angle\beta_2$ quantifies the pattern of the linear polarization spiral in near-horizon images, with $\beta_2=0$ corresponding to radial polarization and $\beta_2=\pm180$ corresponding to azimuthal polarization vectors. The value of $\angle\beta_2$ probes the pitch angle of the underlying magnetic field in the 230 GHz emission region, which is strongly influenced by the black hole spin. More rapidly spinning black holes more efficiently wind up their magnetic fields in powering \citet{BZ} electromagnetic outflows, producing more toroidal fields in the EHT emission region and thus more radial EVPA patterns. In addition, the sign of $\angle\beta_2$ encodes the relative direction of the toroidal and poloidal magnetic fields. For a black hole spin vector oriented $163\deg$ to the line of sight \citep{Mertens2016,PaperV}, negative values of $\angle\beta_2$ (corresponding to right-handed EVPA spiral) indicate magnetic field lines swept back by the black hole spin and thus capable of powering an energy-extracting BZ outflow \citep{Chael23}.

The values of $\angle\beta_2$ in our radiative simulations plotted in \autoref{fig:polstats} reproduce the observed qualitative trend with spin first reported in \citet{Palumbo20}.
Notably, the polarization spiral in the $a_*=-0.1$ simulation \texttt{am1\_radk} has the opposite sense to the rest of the simulations, with  $\angle\beta_2{\approx}160\deg$ when the observer inclination $\theta_{\rm o}=163\deg$. This discrepancy indicates that the magnetic field pitch angle in the EHT emission region is set by the large-scale retrograde accretion flow instead of the black hole spin direction for the weakly spinning $a_*=-0.1$ black hole, just like the asymmetry position angle in \autoref{fig:Istats}.

In the BZ monopole model, the ratio of the Kerr-Schild poloidal to radial magnetic field strength is proportional to $\Omega_{\rm H}$, the angular frequency of the event horizon \citep{BZ,McKinney04}. We find that we can fit a simple two-parameter functional form for $\angle\beta_2(\Omega_{\rm H})$ to the measured $\angle\beta_2$ from the 230 GHz images from the radiative simulations:
\begin{equation}
\label{eq:fit}
\left[\angle\beta_2\right]_{\rm fit} = 2\arctan\left[\frac{-C_0}{\left|\Omega_{\rm H}\right|}\right] + C_1.
\end{equation}
In \autoref{eq:fit}, the parameter $C_0$ encodes how rapidly fieldlines are wound up with increasing black hole spin, while $C_1$ provides an overall constant shift in the zero-spin value of $\angle\beta_2$. We find that this model fits the simulation results well when applied separately to the prograde and retrograde simulations. For the retrograde simulation fit, we find $C_0=0.26,C_1=-26\deg$ and for the prograde simulations we find $C_0=0.11,C_1=17\deg$.
The discrepancy in values of $\angle\beta_2$ and the fitting function results for the prograde and retrograde simulations indicates that the initial condition of the retrograde accretion disc reduces both the rate at which the pitch angle of the field structure increases with spin and the zero value of $\angle\beta_2$.
As a result, a retrograde simulation at a given absolute value of black hole spin will have an overall more toroidal EVPA pattern than its prograde counterpart.

We note again that the average polarization/magnetic field pitch angle encoded in $\angle\beta_2$ is strongly dependent on black hole spin. If we ignore the fact that none of our simulations satisfy EHT constraints on $\langle|m|\rangle$, comparing EHT measurements of $\angle\beta_2$ to our simulations selects a retrograde or very modestly prograde black hole spin, $a_*\in\left[-0.7,0.2\right]$ for M87*.
This inference and the trend for $\angle\beta_2$ presented for the radiative simulations in \autoref{fig:polstats} is consistent with results from our non-radiative simulations, with only small shifts in $\angle\beta_2$ on the order of ${\approx}10\deg$ from Faraday rotation when changing $R_{\rm low}$, $R_{\rm high}$ (see also \citealt{Chael24}, Figure 3 for $\angle\beta_2$ values from the MAD simulations of \citet{Narayan22}). As in \citet{PaperVIII}, we find that raytracing the single-fluid GRMHD simulations with larger values of $R_{\rm high}=80-160$ is necessary to reduce $\langle|m|\rangle$ to within the measured EHT range.

\section{Discussion and Conclusions}
\label{sec:discussion}

In this paper we present the first systematic survey of two-temperature, radiative GRMHD simulations of M87* with mass accretion rates scaled to produce the flux density observed by the EHT at 230 GHz \citep{PaperIV}. We ran eleven 2TGRRMHD simulations in the code \texttt{KORAL} with electrons heated by the \citet{Kawazura19} sub-grid prescription derived from gyrokinetic simulations of plasma turbulence. We find that our simulations have accretion rates $\dot{M}=(0.5-1.5)\times10^{-6}\dot{M}_{\rm Edd}$ and radiative efficiencies $\epsilon_{\rm rad}=3-35\%$, consistent with results from non-radiative simulation fits to EHT observations in \citep{PaperVIII}. We observe that the trends of bulk properties of our simulations with spin are not affected by the addition of radiation at these low radiative efficiencies; in particular, the radiative and non-radiative simulations produce near-identical results for the average horizon magnetic flux $\phi_{\rm BH}$, outflow efficiency $\eta$, and spindown parameter $s$.

The ion and electron temperature profiles of our simulations are relatively independent of the black hole spin, with the electron temperature being ${\approx}5$ times cooler than the ion temperature in the accretion disc. The disc electrons in our simulations are moderately relativistic, and the combined electron-ion fluid has an effective adiabatic index $13/9<\adi<5/3$, while the adiabatic index in the hot jet region is lower, $4/3<\adi<13/9$. We find that the larger value of the adiabatic index in the accretion disc affects global structures in the simulation when compared to our fiducial single-fluid GRMHD simulations with fixed $\adi=13/9$; in particular, the disc scale height in our radiative simulations is larger than in their non-radiative counterparts, and the jet width is correspondingly smaller.

Our simulations produce values of the temperature ratio $R=T_{\rm i}/T_{\rm e}$ that are not tightly correlated with $\beta_{\rm gas}$; by contrast, the standard approach from post-processing modelling of $R$ in single-fluid GRMHD simulations uses a one-to-one function $R(\beta)$.
When fitting the \citet{Moscibrodzka16} model $R(\beta)$ to our simulation results, we find an effective $R_{\rm low}\approx2$ and $R_{\rm high}\approx 5$. Our results for the electron temperature are consistent with results from 2D MAD simulations run with the \citet{Howes10} heating prescription run at comparable accretion rates in \citet{Dihingia20,Dihingia23}.

230 GHz images from our simulations reproduce the total-intensity properties of M87* on event horizon scales observed by the EHT in 2017 \citep{PaperIV} and 2018 \citep{EHT2024M87}. Our 2TGRRMHD simulations also reproduce the same observed trend of the polarization's second Fourier mode phase $\angle\beta_2$ with spin that is observed in single-fluid MAD simulations \citep{Palumbo20} and which directly probes the dragging of magnetic field lines by black hole spin \citep{Chael24}.
However, our simulations produce linear polarization fractions at the resolution of the EHT beam that are several times too large when compared to observations \citep{PaperVII}. These large polarization fractions are also seen in single-fluid MAD GRMHD simulations of M87* raytraced with the \citet{Moscibrodzka16} prescription and small values of the high-$\beta$ ion-to-electron temperature ratio, $R_{\rm high}\lsim10$ \citep{PaperVIII}. In strongly-magnetized MAD GRMHD simulations, cooler electron temperatures (large effective values of $R_{\rm high}\approx80-160$) are necessary to produce sufficient internal Faraday rotation to de-polarize the emission to the observed $\langle |m|\rangle \lsim 10\%$ on $20\,\mu$as scales \citep{PaperVIII,PaperIX}.

The high electron temperatures and correspondingly high polarization fractions in our 2TGRRMHD simulation suite poses a challenge to the interpretation of EHT polarization results. Is it possible to obtain higher ion-to-electron temperature ratio that match values preferred by post-processing modelling of EHT data with physically based models of plasma heating?
First, it should be noted that 2TGRRMHD simulations are known to not perfectly identify grid-scale dissipation; their algorithm for calculating the grid-scale viscous heating rate $q^{\rm v}$ produces artificial heating or even negative heating in certain regions \citep{Ressler17,Sadowski17}.
The \texttt{KORAL} code attempts to counter any trend toward excess heating by correcting $q^{\rm v}$ to account for additional entropy generated by mixing finite-sized fluid parcels (\citealt{Sadowski17}, section 3.1), and recent tests by \citet{Moscibrodzka24} have not found any dependence of the global species temperatures or simulated images on resolution, as might be expected if finite numerical resolution was introducing excess heating.
However, to ensure the numerical method is not responsible for the low $R$ and high polarization values observed here, it would be useful to reproduce the results of our simulation suite with a different numerical implementation.
Ideally, alternate algorithms for identifying $q^{\rm v}$ in simulations should be developed to further test these results.

Another source of uncertainty in comparing GRMHD simulations to images of M87* is the treatment of emission from the magnetized jet region. In this work, we adopt a ``sigma cut-off'' of $\sigma_{\rm cut}=25$, zeroing out 230 GHz emission from all regions with a higher magnetization. A more common practice in the community is to set $\sigma_{\rm cut}=1$, treating emission from all regions where the magnetic field dominates the energy budget as suspect. We find in our simulations between $30-50\%$ of the median 230 GHz flux density comes from the highly magnetized region between $1<\sigma_{\rm i}<25$; we choose to include emission from this region because $\sigma_{\rm i}\approx 1$ gas represents a significant volume of the near-horizon accretion inflow and outflow and contributes meaningfully to the observed emission in MAD systems, while being sufficiently far from the simulation's ceiling value $\sigma_{\rm max}=100$ to avoid obvious artefacts from density floors \citep{Chael19,Chael24}.
However, adopting a lower value of $\sigma_{\rm cut}$ would result in requiring correspondingly higher values of the accretion rate $\dot{M}$ to reproduce the observed 230 GHz flux density $F_{230}$, resulting in somewhat more efficient electron cooling and more Faraday depth, though a $30-50\%$ increase in accretion rate is likely not sufficient to bring our results for $\langle |m|\rangle$ in line with 230 GHz observations.
In the next paper in this series, we will run a detailed comparison of a single two-temperature simulation with accretion rates chosen to reproduce the observed 230 GHz flux density under different $\sigma_{\rm cut}$ assumptions.

While the \texttt{K19} heating prescription seems to produce electrons that are too hot in MAD simulations to satisfy observational constraints, it
may be that another sub-grid heating model exists that supplies a lower $\delta_{\rm e}$ in the 230 GHz emitting region and thus results in cooler electrons and lower polarization from Faraday depolarization.
The logical next step is to compare the simulations presented here with an analogous set of simulations heated by magnetic reconnection using the \citet{Rowan19} (\texttt{R19}) prescription. The \texttt{R19} prescription provides more heat to electrons at high $\beta$, but in the highly magnetized regions of MAD discs where most 230 GHz emission is produced, \texttt{R19} may provide lower values of $\delta_{\rm e}$ than the \texttt{K19} model used in this paper \citep{Chael19}.
Previous studies have found that \texttt{R19} is preferred to a turbulent heating model in reproducing certain features of M87*'s total intensity jet emission \citep{Chael19} and Sgr A*'s total fractional polarization and variability \citep{Chael18, Dexter20}, though Sgr A* is significantly more polarized than M87* on EHT scales \citep{PaperVII}. However, two recent two-temperature radiative simulation studies targeted at Sgr A* \citep{Salas24} and at X-ray binary accretion over a range of accretion rates \citep{Liska24} using the \citet{Rowan17} heating prescription\footnote{\citet{Rowan19} generalized the Particle-in-Cell magnetic reconnection simulations of \citet{Rowan17} to non-zero guide field; they provide different, but qualitatively similar, fitting functions in the zero-guide field case.} both found modest values of $R\approx5-10$, suggesting that a simple change of heating model may not produce sufficiently cold electrons to depolarize M87*'s 230 GHz images.
In the next paper in this series we will investigate if the \texttt{R19} reconnection model is more successful in reproducing the observed low horizon-scale polarization fraction in M87*.

In addition to the original \texttt{K19} heating prescription from Alfv\'enic turbulence, recent work by \citet{Kawazura20} has examined the role of compressive driving of a turbulent cascade in altering the resulting electron heating fraction at small scales. \citet{Satapathy23,Satapathy24} have explored the implications of the \citet{Kawazura20} model in semi-analytic models and shearing-box simulations, and have found that a larger degree of compressive driving lowers the effective $\delta_{\rm e}$ from the \citet{Kawazura19} \texttt{K19} result, with a predicted range of effective temperature ratios $R\approx5-40$ for $\beta_{\rm gas}\approx1-10$. Global 2TGRRMHD simulations including an effective subgrid model for turbulent heating modified by compressive modes following \citet{Satapathy24} are necessary to establish if this change is sufficient to produce polarization fractions for M87* within the measured range for electrons heated by plasma turbulence.

If no plasma heating prescription $\delta_{\rm e}$ can be found that produces sufficiently cold electrons and sufficiently large depolarization in MAD M87* models, it may imply that other changes to the accretion flow structure independent of the electron temperature modelling should be considered. Turbulent field ``SANE'' models are naturally more depolarized than MAD models at all values of the \citet{Moscibrodzka16} parameters $R_{\rm high}$, $R_{\rm low}$ due to their more turbulent magnetic fields \citep{PaperVIII}. SANE simulations generally do not satisfy EHT constraints on $\angle\beta_2$ and can in some cases be too depolarized to satisfy the observed lower bound on $\langle|m|\rangle$, and EHT simulation scoring largely rules them out as models of M87* when compared to MAD simulations \citep{PaperVIII}. However, if electrons in two-temperature simulations cannot be made sufficiently cool it may be that simulations with a more intermediate level of $\phi_{\rm BH}$ higher than the typical value $\phi_{\rm BH}\lsim5$ but below the MAD saturation of $\phi_{\rm BH}\approx50$ may produce more depolarized emission while satisfying EHT constraints on $\beta_2$.
More broadly, future investigations should explore the effect on the polarization of physics and initial conditions not explored in the standard assumptions of aligned, two-temperature accretion such as
disc tilt \citep{White19,Chatterjee20},
large-scale plasma feeding \citep{Guo24},
anisotropic electron distributions \citep{Galishnikova2023}, and
non-zero helium fractions \citep{Wong2022}.

Previous comparisons of 2TGRRMHD simulations to total intensity and pre-EHT observations of Sgr A* \citep[e.g][]{Chael18} and M87* \citep[e.g][]{Chael19} found only weak preferences for different electron heating mechanisms based on the observational data; prior to the first EHT polarized images, constraining the microscopic plasma heating processes in Sgr A* and M87* with observations did not seem feasible.
The advent of robust polarized images from the EHT have changed this paradigm.
The results of this first large survey of 2TGRRMHD simulations of M87* indicate that
polarized horizon-scale images can strongly constrain models of collisionless plasma heating in hot accretion flows just outside supermassive black holes.

\section*{Acknowledgements}
We thank George Wong, Eliot Quataert, Monika Mo\'scibrodzka, and Matthew Liska for useful conversations and suggestions. This work used the Stampede2 and Stampede3 resources at TACC through allocation TG-AST190053 from the Extreme Science and Engineering Discovery Environment (XSEDE), which were supported by National Science Foundation grant numbers \#1548562, \#2320757.
The authors acknowledge the Texas Advanced Computing Center (TACC) at The University of Texas at Austin for providing {HPC, visualization, or storage} resources that have contributed to the research results reported within this paper. URL: http://www.tacc.utexas.edu.
Computations in this paper were run on the FASRC cluster supported by the FAS Division of Science Research Computing Group at Harvard University

\section*{Data Availability}
The data from the simulations reported here are available on request. The \texttt{KORAL} code is available at \url{https://github.com/achael/koral_lite}.




\bibliographystyle{mnras}
\bibliography{allrefs} 




\bsp	
\label{lastpage}

\end{document}